\documentclass{emulateapj}
\pdfoutput=1

\usepackage{bm}
\usepackage{times}
\usepackage{verbatim}
\usepackage{graphicx}
\usepackage{graphics,epsfig}
\usepackage{theorem}
\usepackage{makeidx}
\usepackage{amsmath}
\usepackage{epic}
\usepackage{amscd}
\usepackage{bbm}
\usepackage{xy}
\usepackage{amssymb}
\usepackage{epsfig}

\usepackage{amsbsy}

\def\bs{\boldsymbol}
\def\gsim{\;\rlap{\lower 2.5pt
\hbox{$\sim$}}\raise 1.5pt\hbox{$>$}\;}
\def\lsim{\;\rlap{\lower 2.5pt f
\hbox{$\sim$}}\raise 1.5pt\hbox{$<$}\;}

\newcommand{\gtsimeq}{\raisebox{-0.6ex}{$\, \stackrel{\raisebox{-.2ex}%
{$\textstyle >$}}{\sim}\,$}}
\newcommand{\ltsimeq}{\raisebox{-0.6ex}{$\, \stackrel{\raisebox{-.2ex}%
{$\textstyle <$}}{\sim}\,$}}

\begin{document}

\title{Mixing of  Clumpy Supernova Ejecta into Molecular Clouds}

\author{Liubin Pan, Steven J.\ Desch, Evan Scannapieco, \& F.~X.~Timmes}

\affil{School of Earth and Space Exploration, Arizona State University,  P.~O.~Box 871404, Tempe, AZ, 85287-1404}

\begin{abstract}

Several lines of evidence, from isotopic analyses of meteorites to studies of the Sun's
elemental and isotopic composition, indicate that the solar system was contaminated
early in its evolution by ejecta from a nearby supernova.
Previous models have invoked supernova material being injected into an extant protoplanetary
disk, or isotropically expanding ejecta sweeping over a distant ($> 10$ pc)
cloud core, simultaneously enriching it and triggering its collapse.
Here we consider a new astrophysical setting: the injection of clumpy supernova ejecta, as 
observed in the Cassiopeia A supernova remnant, into the molecular gas at the periphery of an 
H {\sc ii} region created by the supernova's progenitor star.  
To track these interactions we have conducted a suite of high-resolution ($1500^3$ effective) 
three-dimensional numerical hydrodynamic simulations that follow the evolution of individual 
clumps as they move into molecular gas. 
Even at these high resolutions, our simulations do not quite achieve numerical convergence, 
due to the challenge of properly resolving the small-scale mixing of ejecta and molecular gas, 
although they do allow some robust conclusions to be drawn.
Isotropically exploding ejecta do not penetrate into the molecular cloud or mix with it, 
but, if cooling is properly accounted for, clumpy ejecta penetrate to distances 
$\sim 10^{18} \, {\rm cm}$ and mix effectively with large regions of star-forming 
molecular gas. 
In fact, the $\sim 2 \, M_{\odot}$ of high-metallicity ejecta from a single core-collapse 
supernova is likely to mix with $\sim 2 \times 10^{4} \, M_{\odot}$ of molecular gas material 
as it is collapsing.
Thus all stars forming late ($\approx 5 \, {\rm Myr}$) in the evolution of an H {\sc ii} region
may be contaminated by supernova ejecta at the level $\sim 10^{-4}$.
This level of contamination is consistent with the abundances of short-lived radionuclides
and possibly some stable isotopic shifts in the early solar system, and is potentially 
consistent with the observed variability in stellar elemental abundances.
Supernova contamination of forming planetary systems may be a common, universal process. 

\end{abstract} 


\section{Introduction}

\subsection{Solar System Contamination by Supernova Material}

Many lines of evidence indicate that our solar system was contaminated during its formation 
by material from a nearby core-collapse supernova. Isotopic analyses of meteorites 
reveal both evidence for the one-time presence of short-lived radionuclides (SLRs) as 
well as stable element isotopic anomalies suggestive of supernova ejecta.   
Furthermore, the Sun's elemental and even its isotopic composition point to 
contamination from a supernova.

Traditionally, the strongest arguments for supernova contamination 
come from isotopic analyses of the decay products of radioactive isotopes in meteorites.
By observing correlations between excesses of the daughter isotope with the elemental abundance 
of the parent, it is inferred that the solar nebula contained several SLRs with half lives 
$< 10 \, {\rm Myr}$, including ${}^{36}{\rm Cl}$, ${}^{10}{\rm Be}$, and most importantly, 
${}^{26}{\rm Al}$, and ${}^{60}{\rm Fe}$ (Wadhwa et al.\ 2007).
Even before it was discovered, Cameron (1962) suggested that the presence of ${}^{26}{\rm Al}$ 
in the early solar system would imply injection from a nearby supernova. 
Since its discovery (Lee et al.\ 1976), alternative sources of ${}^{26}{\rm Al}$  
have been suggested, including production by irradiation by energetic particles within the 
solar nebula (Lee et al.\ 1998; Gounelle et al.\ 2001, 2006).
These models encounter a number of difficulties, however (Desch et al.\ 2010),
and an external nucleosynthetic source is usually invoked for this isotope
(Huss et al.\ 2009; Wadhwa et al.\ 2007; Makide et al.\ 2011; Boss 2012).

More recently, the existence of ${}^{60}{\rm Fe}$ in the solar nebula at a level 
${}^{60}{\rm Fe} / {}^{56}{\rm Fe} \sim 3 \times 10^{-7}$ was reported
by Tachibana \& Huss (2003).  
This would definitively indicate injection of material from a nearby supernova into the Sun's
molecular cloud or protoplanetary disk, as no other plausible sources exist for this neutron-rich 
isotope (Leya et al.\ 2003; Wadhwa et al.\ 2007).
On the other hand, the widespread  existence of ${}^{60}{\rm Fe}$ in the solar nebula at 
these levels has been called into question, although its existence at lower levels, 
${}^{60}{\rm Fe} / {}^{56}{\rm Fe} \sim 1 \times 10^{-8}$ appears to be robust 
(Telus et al.\ 2012; Quitte et al.\ 2010; Spivak-Birndorf et al.\ 2011).
Even at ${}^{60}{\rm Fe} / {}^{56}{\rm Fe} \sim 1 \times 10^{-8}$, the existence of 
${}^{60}{\rm Fe}$ probably demands a late input from a supernova (Jacobsen 2005; Huss et al.\ 2009).
Thus, while the evidence from meteoritic SLRs is not quite as clear-cut as previously 
thought, the consensus view remains that ${}^{60}{\rm Fe}$, ${}^{26}{\rm Al}$, and other 
SLRs were injected by a supernova.

Furthermore, the SLR measurements in meteorites also suggest this contamination 
occurred early in the solar system's evolution (Wadhwa et al.\ 2007; Huss et al.\ 2009).
This is because high levels of ${}^{26}{\rm Al}$ (at an initial abundance 
${}^{26}{\rm Al} / {}^{27}{\rm Al}$ $\approx 5 \times 10^{-5}$) are commonly inferred for 
calcium-rich, aluminum-rich  inclusions (CAIs) in meteorites at the time they formed 
(MacPherson et al.\ 1995). 
CAIs are composed of minerals that condense from a solar composition gas at very high
temperatures, $> 1700 \, {\rm K}$ (Ebel \& Grossman 2000), meaning that they formed in 
a hot solar nebula.
Such temperatures require high mass accretion rates through the protoplanetary disk
$\dot{M} > 10^{-6} \, M_{\odot} \, {\rm yr}^{-1}$ that cannot be maintained 
for more than $\sim 10^{5} \, {\rm yr}$ (e.g., Lesniak \& Desch 2011). 
This timeframe is consistent with the finding by Larsen et al.\ (2011) that the 
initial ${}^{26}{\rm Al} / {}^{27}{\rm Al}$ ratio in CAIs is uniform and suggestive 
of ${}^{26}{\rm Al}$-bearing CAIs forming from a homogenized reservoir all within 
$< 3 \times 10^{5} \, {\rm yr}$ of each other (Makide et al.\ 2011). 
In fact, this timescale is nearly as short as the expected free-fall timescale on which 
material is believed to collapse from the molecular cloud, and it appears quite likely
that ${}^{26}{\rm Al}$ was injected at some point {\it during} the collapse process
(Thrane et al.\ 2008; Makide et al.\ 2011).
Injection and incomplete homogenization would also explain the existence of rare
CAIs called FUN inclusions (Fractionation and Unknown Nuclear effects), for which
strong upper limits on initial ${}^{26}{\rm Al}$ exist, as low as 
${}^{26}{\rm Al} / {}^{27}{\rm Al} < 10^{-8}$ (Fahey et al.\ 1987), at least 
for some of these objects. 
Presumably these CAIs formed early, from material not yet contaminated by mixing of 
injected supernova material (Sahijpal \& Goswami 1998). 
The weight of evidence is that injection of ${}^{26}{\rm Al}$-bearing supernova 
material happened very early in the solar system's evolution, probably in the first 
1 Myr. 

Strong meteoritic evidence for supernova injection is also provided by 
stable isotope anomalies.
Variations in ${}^{54}{\rm Cr}$ among planetary materials argue strongly
for a heterogeneous distribution of this isotope within the solar nebula
(Podosek et al.\ 1997; Rotaru et al.\ 1992; Trinquier et al.\ 2007).
The carrier of this anomaly recently has been discovered to be small 
($\sim 100 \, {\rm nm}$) spinel (${\rm MgAl}_{2}{\rm O}_{4}$) presolar 
grains with ${}^{54}{\rm Cr} / {}^{52}{\rm Cr}$ ratios greater than 3 
(Dauphas et al.\ 2010) or more (Qin et al.\ 2011; Nittler et al.\ 2012) times the solar 
value.
Qin et al.\ (2011) argue these formed from material from the O/Ne and O/C 
burning zones of a type II supernova.

Other stable isotopes anomalies appear to correlate with ${}^{54}{\rm Cr}$, 
including ${}^{62}{\rm Ni}$ (Regelous et al.\ 2008) and ${}^{46}{\rm Ti}$ 
and ${}^{50}{\rm Ti}$ (Trinquier et al.\ 2009), which
Qin et al.\ (2011) argue are also consistent with an origin in the O/Ne or O/C 
zones of a type II supernova.
Interestingly, Larsen et al.\ (2011) have presented evidence for 
heterogeneous ${}^{26}{\rm Mg}$ anomalies (from decay of ${}^{26}{\rm Al}$) 
that correlate with the ${}^{54}{\rm Cr}$ anomalies, which would strongly 
imply that the source of ${}^{26}{\rm Al}$ in the solar nebula was associated 
with the nanospinels that introduced the ${}^{54}{\rm Cr}.$ 
In addition, Ranen \& Jacobsen (2006) inferred late contributions from a 
nucleosynthetic source from variations in Ba isotopes, and Dauphas et al.\ (2002) 
inferred the same from variations in Mo isotopes. 

These stable isotope anomalies, manifested as differences in isotopic ratios
between different planetary materials, represent (late) additions of material 
that did not mix well in the solar nebula.
There are also stable isotopes which appear well mixed but manifest themselves
as differences in isotopic ratios between planetary materials and the predictions
of Galactic chemical evolution. 
As emphasized by Clayton (2003), the isotopic ratios of Si in meteorites and 
planetary materials in the solar system are difficult to reconcile with the 
isotopic ratios in ``mainstream" SiC presolar grains. 
These grains seem to show greater contributions from secondary isotopes
(${}^{29}{\rm Si}$ and ${}^{30}{\rm Si}$), relative to the primary isotope 
${}^{28}{\rm Si}$, than solar system materials, despite the fact that they predate 
the solar system and  sample material that has seen less Galactic chemical evolution 
(Clayton \& Timmes 1997; Alexander \& Nittler 1999; Zinner 1998).
Contamination of the solar system by ${}^{28}{\rm Si}$-rich supernova material 
has been invoked to explain this discrepancy (Alexander \& Nittler 1999).
In a similar way, Young et al.\ (2011) have considered the oxygen isotopic composition 
of the solar system in a Galactic context, comparing it to gas around protostars.
They infer that the solar system was enriched in ${}^{18}{\rm O}$ (and / or depleted in 
${}^{17}{\rm O}$), relative to ${}^{16}{\rm O}$, by about 30\%.
They also argue for mixing of material with ejecta from a core-collapse supernova. 

Going beyond the strong evidence for supernova contamination of meteorites and 
planetary materials, there is growing evidence for contamination of the Sun itself.  
Recent {\it Genesis} measurements of isotopic ratios in the solar wind appear
to confirm that the Sun's oxygen isotopic ratio matches that of CAIs in meteorites 
(McKeegan et al.\ 2011), meaning that if the meteorites differ isotopically from the 
Galactic average, then so does the Sun.
Also, it has long been recognized that the Sun's metallicity is anomalously high compared 
to G dwarfs formed at the same time and galactocentric distance (Edvardsson et al.\ 1993), 
and it has even been suggested that the Sun formed at 6.6 kpc, in order to explain its 
elevated [Fe/H] (Wielen et al.\ 1996). 
An alternative explanation is that stars forming at the same place and time may receive 
considerably different contributions of supernova material (Reeves 1978).
The Sun's [Fe/H] might appear anomalously high if it received a significant amount
of supernova material.

A prediction of this scenario is that stars would exhibit variations in [Fe/H] and other
elemental ratios, because of the presumably stochastic nature of supernova contamination.
Observational support for elemental variations was sought by Cunha \& Lambert (1994) 
and Cunha et al.\ (1998), who found up to a factor of 2 variations in elemental
ratios in O and Si but not Fe, C and N among newly formed B, F and G stars of the same age 
and subgroup in the Orion star-forming region.
The variability of O and Si, which are primary products of core-collapse supernovae, but
not in C and N, which come predominantly from sources other than core-collapse supernovae,
was taken as strong evidence for contamination from nearby supernovae. 

Unfortunately, subsequent work has not confirmed such high degrees of variability among 
Orion stars, (D'Orazi et al.\ 2009; Takeda et al.\ 2010; Sim\'{o}n-Di\'{a}z 2010; Nieva 
\& Sim\'{o}n-Di\'{a}z 2011). 
Intriguingly, though, among stars known by radial velocity surveys to host planets, 
the ratios of abundant elements like C, O, Si and Fe appear to vary by factors of 2 in 
their stellar atmospheres (Bond et al.\ 2008; Pagano et al.\ 2010).
Supernova injection into the molecular cloud from which protostars are forming remains 
a plausible mechanism for these variations, and may contribute to the abundances observed in 
planet-hosting stars. 

In summary, the preponderance of the evidence from studies of
SLRs and stable isotope anomalies in meteorites, comparisons of
stable isotopic ratios in the solar system with those in presolar grains and
interstellar gas, and measurements of the elemental variations of planet-hosting stars
stars all point to a single scenario.  
Supernova ejecta contaminated the Sun, likely very early in the solar system's
evolution,  and similar contamination is likely to be a common occurrence in the formation 
of Sun-like stars. 

\subsection{Sources of Supernova Contamination} 

Various models have been proposed for how a newly forming solar system could be 
contaminated with supernova material  during either in the  early stages of collapse, 
or soon after the protoplanetary disk has formed. 
Cameron \& Truran (1977) suggested that the Sun's molecular cloud core was both
contaminated by supernova material {\it and} simultaneously triggered by the 
supernova shock to collapse. 
Increasingly sophisticated numerical models have simulated the 
interaction of supernova ejecta with a marginally stable molecular cloud core,
showing that the ejecta simultaneously can trigger collapse of 
the cloud core and inject supernova material into the collapsing gas, provided 
the ejecta have been slowed to speeds $5 - 70 \, {\rm km} \, {\rm s}^{-1}$ 
(Boss 1995; Foster \& Boss 1996, 1997; Boss \& Foster 1998; Vanhala \& Cameron 
1998; Boss \& Vanhala 2000; Vanhala \& Boss 2002; Boss et al.\ 2008, 2010; Boss 
\& Keiser 2010).
This last point is crucial, since higher speeds tend to shred apart the cloud 
core rather than initiate its collapse. The need to slow the ejecta 
from initial velocities $> 2000 \, {\rm km} \, {\rm s}^{-1}$ demands that 
several parsecs of gas must lie between the supernova and the
cloud core. Such molecular gas is observed to lie at the periphery of H {\sc ii} 
regions in which massive stars evolve and go supernova, but the rest of the scenario is difficult 
to test observationally, because the cloud cores would be deeply embedded several
parsecs deep within the molecular clouds.

Supernova injection into molecular clouds was explored in a different context by Gounelle et al.\ (2009).
In their model, multiple supernovae in a stellar cluster sequentially condense the ambient low-density
interstellar gas into molecular clouds, and the ejecta material is assumed to mix into the
molecular gas simultaneously. As a result of this sequential enrichment, stars of the next generation 
forming from these molecular clouds would contain the products of multiple supernovae.


Injection of supernova material into a protoplanetary disk was considered 
analytically by Chevalier (2000), and numerically by Ouellette et al.\ (2005, 2007, 2009, 2010).
These authors noted that protoplanetary disks are commonly found in high-mass
star-forming regions, near massive stars that will quickly evolve off the main
sequence and explode as supernovae. 
Ouellette et al.\ (2010) found that injection of supernova material into a 
protoplanetary disk, at levels high enough to explain the abundances of SLRs
like ${}^{26}{\rm Al}$, was possible, provided these species resided in large 
(radii $> 0.1 \, \mu{\rm m}$) dust grains, to avoid flowing around the disk.
Gounelle \& Meibom (2009) and Williams \& Gaidos (2009) noted that the disk is very 
likely to have already evolved several Myr, or to be many parsecs away, at the time of 
the explosion, raising doubts that injection into a protoplanetary disk can explain the 
abundances of SLRs.
Ouellette et al.\ (2010) countered that a combination of triggered formation
and clumpy supernova ejecta may yet satisfy the constraints, and work on this
model is ongoing.
An important feature of this model that may distinguish it from alternatives is
that significant amounts of supernova ejecta do not enter the star, just the disk
material.

Here we study a third alternative, based on the observation that star formation 
occurs in the molecular gas at the edges of H {\sc ii} regions, quite probably 
triggered by the ionization fronts and associated shocks driven by the massive stars 
at the center of the H {\sc ii} region (Hester et al.\ 2004; Hester \& Desch 2005; 
Snider et al.\ 2009).
Supernova ejecta from the explosion of a massive star will generally occur at the
center of an H {\sc ii} region, and will generally contaminate this peripheral gas.  
If supernova ejecta could emplace themselves in the molecular gas as it collapses due 
to compression either from a D-type ionization front or a supernova shock, then supernova 
contamination of protostars would indeed be a common process. 
Thus we are motivated to study a mechanism by which supernova material may be deposited
directly into forming solar systems:  the injection of dense clumps of innermost supernova 
material such as those observed in SN 1987A and the Cassiopeia A (Cas A) supernova remnant.
By means of high-resolution 3D numerical simulations, we consider how such highly-enriched dense
knots enter and mix with the gas of a nearby molecular cloud, at the periphery
of the H {\sc ii} region in which the supernova progenitor resided.

Many interesting numerical studies of the interaction of cold, overdense clumps moving 
through a hot, lower density medium have been undertaken in other astrophysical contexts. 
Klein et al.\  (1994) studied the evolution of nonradiative clouds propagating through the 
general interstellar medium (ISM), showing that if the cloud velocity is much greater than 
its sound speed, it will be disrupted on a ``cloud crushing" timescale given by the time for 
the shock to cross the cloud interior.  
Subsequent ISM-scale studies showed that  magnetic fields (e.g.\  Mac Low et al.\ 1994) and 
radiative cooling that operates above the initial cloud temperature (Fragile et al.\ 2005) 
were only able to delay this disruption by 1-2 cloud crushing times.   
However, if shock interactions are able to efficiently catalyze coolants that radiate 
efficiently below the initial cloud temperature, the cloud will collapse, a process that may 
lead to triggered star formation on galactic scales (e.g., Fragile et al. 2004;  
Gray \& Scannapieco 2010).   

Even if cooling is efficient only above the initial cloud temperature, clumps can be maintained 
for long timescales if they move through the medium faster than the exterior sound speed, because 
of a bow shock that forms in front of them.   
Analogous cases that have been modeled include comet Shoemaker-Levy 9 plunging through 
Jupiter's atmosphere (Mac Low \& Zahnle 1994), clouds interacting with galaxy outflows 
(Cooper et al.\ 2009), high-velocity clouds orbiting the Milky Way (Kwak et al.\ 2011), and  
``bullets" of ejecta from stellar outflows (Poludnenko et al.\ 2004),  proto-planetary nebulae 
(Dennis et al.\ 2008), and supernovae (Raga et al.\ 2007)  moving through the ionized ISM.  
The simulations presented in this paper also lie in this supersonic regime, but invoke a 
very different set of parameters from these previous studies. 
Our simulations are the first to study the interaction of supernova bullets with the molecular
gas at the periphery of an H {\sc ii} region.

The structure of this paper is as follows.
In \S 2 we discuss the astrophysical context in which supernovae often take place, in an
H {\sc ii} region.  We also discuss the evidence that a substantial portion of supernova 
ejecta explodes in the form of dense clumps.
In \S 3 we outline the numerical methods by which we model the interaction of these clumps 
with the surrounding molecular cloud.
In \S 4 we present the results of a parameter study designed to test the extent to which 
supernova material can penetrate into a molecular cloud and mix with molecular gas, as a 
function of clump velocity, mass, density, and other parameters. 
In \S 5 we discuss the implications of these results for the abundances of short-lived 
radionuclides and stable isotope anomalies in the early solar system, elemental variations 
among stars formed in the same cluster, and galactic enrichment in general.

\section{Astrophysical Context}

\subsection{Star Formation in H II Regions} 

The environment of star formation has a large bearing on the sequence of 
events surrounding the explosion of a supernova and its injection into a 
forming solar system.
First, it is important to recognize that most Sun-like stars form in 
massive star forming regions.
The complete census by Lada \& Lada (2003) of stellar clusters embedded 
in molecular clouds shows that at least 70\%, and probably closer to 90\%,
of {\it all} stars form in embedded clusters, and 90\% of those stars that 
do (i.e., $\approx 81\%$ of all stars) form in a cluster with mass $> 10^2 \, M_{\odot}$.
Clusters can reach masses $\sim 10^6 \, M_{\odot}$ (e.g., the Carina Nebula
has mass $\approx 3 \times 10^{5} \, M_{\odot}$; Preibisch et al.\ 2011). 
The cluster initial mass function (IMF) suggests that the number of clusters 
with mass $M$ scales as $dN/dM \sim M^{-\alpha}$ over the range $10^2 - 10^6 \, M_{\odot}$, 
with $\alpha$ observed to be in the range 1.6 - 1.8 (Elmegreen \& Falgarone 1996).
Assuming similar star formation efficiencies and stellar initial mass functions 
in all clusters, this suggests that of all the stars born in embedded clusters with 
mass $> 10^2 \, M_{\odot}$, 
$\approx 93\%$ (i.e., $75\%$ of all stars) form in clusters with mass 
$> 10^3 \, M_{\odot}$.
This mass cutoff (roughly the size of the Orion Nebula: Hillenbrand et al.\ 2001)
is important because clusters above this size are likely to
contain at least one star with mass $> 40 \, M_{\odot}$ that will rapidly explode as 
a supernova (Adams \& Laughlin 2001).
Thus about three quarters of all Sun-like stars form in a region that will experience
a ``prompt" supernova. 

The time for a star to explode as a supernova depends, of course, on its mass.
Stellar evolution models typically predict that the progenitors of 
core-collapse supernovae will stay on the main sequence for 3-20 Myr: 
progenitors of masses $25 \, M_{\odot}$, $40 \, M_{\odot}$ and $60 \, M_{\odot}$
will explode after about 7 Myr, 5 Myr and 4 Myr, respectively
(Maeder \& Meynet 1989; Schaller et al.\ 1992).
The Orion Nebula, with about 3000 stars, contains one star that will explode
as a supernova within 5 Myr.
Richer clusters are more likely to contain more massive stars that evolve 
faster (Adams \& Laughlin 2001).  
Notably, while fewer than $10\%$ of clusters remain bound for more than 
about 10 Myr (Lada \& Lada 2003), a cluster almost certainly stays intact 
for at least 5 Myr, while significant gas remains.
Thus, half of Sun-like stars form in clusters in which the supernova occurs
in the first 4-5 Myr of the cluster lifetime. 

Star formation appears to continue throughout the evolution of a rich cluster.
Young ($< 10^5$ yr) protostars are often seen even in H {\sc ii} regions even 
several Myr old (Palla \& Stahler 2000; Hester et al.\ 1996, 2004; Healy et al.\ 2004; 
Sugitani et al.\ 2002; Snider 2008; Snider et al.\ 2009; Snider-Finkelstein 2010; 
Getman et al.\ 2007; Reach et al.\ 2009; Choudhury et al.\ 2010; Billot et al.\ 2010; 
Bik et al.\ 2010; Zavagno et al.\ 2010; Beerer et al.\ 2010; Comer\'{o}n \& Schneider 2011).
Many of these authors attribute the late formation of the protostars to triggering by the
advancing ionization fronts launched by the O stars in the cluster, which are the 
progenitors of the supernovae.
We return to this argument in \S 5, as it bears directly on the statistical likelihood
that a {\it newly} formed ($< 1$ Myr old) protostar can be contaminated by supernova 
ejecta.
For now we note that protostars do apparently form throughout the evolution of 
a rich cluster, and that we expect them to be  forming  when the massive stars 
go supernova.

When the most massive star in a cluster goes supernova, the protostars in the process
of forming will lie several parsecs from the supernova. 
Within an H {\sc ii} region, the most massive stars are generally found near 
the center of the spatial distribution of stars (e.g., in the Orion Nebula Cluster: 
Hillenbrand \& Hartmann 1998), where they may have formed, or relaxed dynamically on
very short ($< 10^5$ yr) timescales (Allison et al.\ 2009). 
Protostars, on the other hand, must form from the molecular gas on the periphery of
the H {\sc ii} region. 
The distance of the H {\sc ii} region edge from the massive stars at its center is 
set by the rate of advance of the ionization front carving out the H {\sc ii} region.
The speed of propagation of this front depends not just on the ultraviolet (UV) flux 
from the massive stars but also the density of molecular gas.
Because of uncertainties in physical quantities and the rate of recombinations in the 
ionized gas, it is difficult to predict the speed of an ionization front from first 
principles.
Nevertheless, propagation speeds $0.1- 1.0 {\rm km} \, {\rm s}^{-1}$ are typically
inferred (Osterbrock 1989), both from simulations (Miao et al.\ 2006), and observations
(White et al.\ 1999; Getman et al.\ 2007; Choudhury et al.\ 2010).
Taking $0.4 \, {\rm km} \, {\rm s}^{-1}$ as a typical speed, we infer that by
the time of the first supernova in an H {\sc ii} region, at age 5 Myr, molecular gas 
lies roughly 2 pc from the explosion.
Ejecta travelling at $\approx \, 2000 \, {\rm km} \, {\rm s}^{-1}$ will cross this
distance in only $\approx$ 1000 yr, and will encounter molecular gas in which protostars 
are forming.

\subsection{Isotropic vs. Clumpy Supernova Ejecta} 

The interaction between the molecular gas and the supernova ejecta that collide
with it will depend greatly on the spatial distribution of the ejecta, and especially
the ejecta density.
Numerical simulations of supernova explosions generally show that the outer layers
(the H- and He-burning shells) explode isotropically, but the shells interior to this
are subject to numerous RT and Richtmyer-Meshkov instabilities at 
compositional interfaces.
These instabilities concentrate much of this interior ejecta into dense clumps
(Arnett et al.\ 1989; Fryxell et al.\ 1991; M\"{u}ller et al.\ 1991; Herant \& Benz 1991, 1992; 
Hachisu et al.\ 1991, 1992; Nagataki et al.\ 1998; Kifonidis et al.\ 2003, 2006; Joggerst et al.\ 
2009, 2010; Hammer et al.\ 2010; Ellinger 2011). 
In these simulations, some instabilities at the He/H interface are often seen, but
they are considerably stronger at the He/C and other interfaces. 

Strong evidence for clumpiness exists from observations of nearby supernova remnants.
Ejecta in SN1987A, especially the innermost, Fe-bearing portions, appear clumpy.
The early appearance of gamma rays (Matz et al.\ 1988) is consistent with concentration
of ${}^{56}{\rm Ni}$ into high-velocity clumps (Lucy et al.\ 1988).
Fe emission was far lower than expected for optically thin gas (Haas et al.\ 1990).
Fe {\sc ii} emission disappeared around day 640 (Colgan et al.\ 1994), just as gas emission
became absorbed by dust (Lucy 1988; Colgan et al.\ 1994), and blackbody emission by 
dust arose (Wooden et al.\ 1993).  
The observations of Wooden et al.\ (1993) showed that the dust emission was optically thick, 
even at $30 \, \mu{\rm m}$, strongly implying optically thick clumps. 
Clumpiness is also manifest in Cas A.
In both optical-wavelength {\it HST} images (Fesen et al.\ 2001; Fesen 2005) and in 
high-resolution {\it Chandra Observatory} X-ray images (Hwang et al.\ 2004; Patanude \& 
Fesen 2007), numerous knots of emission are seen, interpreted as clumpy ejecta passing
through the reverse shock (McKee 1974). These ejecta knots are typically 0.2" to 0.4" in 
size, or about $0.5 - 1 \times 10^{16} \, {\rm cm}$ in radius, but may have structure 
at smaller scales.  {\it HST} observations of the nearby Cygnus Loop supernova 
remnant reach a resolution scale of $\ltsimeq 10^{15}$ cm (0.1") (Blair et al.\ 1999). However, 
due to its age, the physical condition in the Cygnus Loop remnant is likely inapplicable to 
the ejecta properties at the early phase we are interested in. On the other hand, 
it is difficult to directly image fine structures in distant supernova remnants, but 
there is no reason to conclude that clumpiness is not a universal process at the early stage of supernova explosions. 

Ouellette et al.\ (2010) showed that the numerical simulations and observations of SN1987A
and Cas A are all consistent with a large fraction of the ejecta mass inside the H 
envelope exploding in the form of homologously expanding clumps.
They argued for $\sim 10^{4}$ clumps, each of mass $\sim 2 \times 10^{-4} \, M_{\odot}$, 
and radii $\approx  1/300$ of the distance from the explosion center, as seen in Cas A.
The volume filling fraction of these clumps is $3.7 \times 10^{-4}$, and if they contain 
most of the mass of the innermost ejecta, they will be a factor $\approx 2700$ denser than 
the average density in an isotropic explosion.
Both the numerical simulations and observations are biased toward the largest clumps, and
it must be understood that smaller clumps are also possible and may be more numerous. 

Strong support for such dense clumps also comes from modeling of dust condensation in supernova
ejecta.
Some presolar grains contain isotopic signatures of condensation from supernova ejecta:
for example, presolar SiC grains of supernova origin  show evidence for large 
excesses of ${}^{44}{\rm Ca}$ resulting from the decay of the neutron-rich isotope 
${}^{44}{\rm Ti}$ 
and some presolar graphite grains
show isotopic evidence for condensation from supernova ejecta (Zinner et al.\ 2007). 
These grains must form at about 1 year after the supernova explosion, after the ejecta 
have expanded and adiabatically cooled, but before the density has dropped too low
for condensation (Kozasa et al.\ 1991, 2009; Nozawa et al.\ 2003, 2010).
Fe grains were observed to form in the SN 1987A remnant after about 600 days 
(Wooden et al.\ 1993).
High ejecta densities are more favorable for dust condensation.

Regarding presolar supernova graphites in particular, Fedkin et al.\ (2010) have shown that 
C-rich supernova ejecta should condense into grains in the sequence TiC, then graphite, then FeSi, 
SiC and metal, {\it unless} the pressures of the ejecta gas exceed $\sim 10^{-5} \, {\rm bar}$, 
in which case graphite condenses after the other phases.
In presolar graphite grains, graphite clearly condensed last (Croat et al.\ 2011), implying 
pressures $> 10^{-5} \, {\rm bar}$ and densities orders of magnitude higher than would 
exist in isotropically expanding ejecta.
For example, if $2 \, M_{\odot}$ of ejecta expands isotropically outward until its temperature
drops below the threshold $\approx 2000$ K necessary for dust condensation, which takes about
1 year (Kosaza et al.\ 1991, 2010), then at a speed of 2000 km s$^{-1}$ it will have expanded 
to a radius 0.002 pc, and its pressure will be $\approx 4 \times 10^{-10}$ bar (Kozasa et al.\ 2010)
to $1 \times 10^{-9}$ bar (Nozawa et al.\ 2003).
Pressures $> 10^{-5} \, {\rm bar}$ require clumps that are overdense by factors $> 10^{4}$ if
the condensation takes place at 1 year, or $\sim 10^{2}$ if the temperature drops below 2000 K
at an earlier, denser stage at a few months.
The condensation sequence of supernova graphites demands that they formed in clumpy ejecta
significantly denser than isotropically exploding ejecta. 

The existence of two ``phases" of supernova ejecta---isotropically exploding outer layers, and
clumpy inner layers---completely changes how ejecta will interact with the molecular cloud.
Assuming as much as $20 \, M_{\odot}$ of supernova material (in the H shell) explodes 
isotropically, producing a shell extending from, say 2.0 pc to 2.4 pc, its mean density will 
be $\approx 6 \times 10^{-23} \, {\rm g} \, {\rm cm}^{-3}$, 
and its surface density will be $\approx 8 \times 10^{-5} \, {\rm g} \, {\rm cm}^{-2}$.
This density is significantly lower than the density in the molecular gas, which for 
$n_{\rm H2} = 10^4 \, {\rm cm}^{-3}$ is $4.7 \times 10^{-20} \, {\rm g} \, {\rm cm}^{-3}$.
In contrast, the $\approx 2 \times 10^{-4} \, M_{\odot}$ clumps modeled by Ouellette 
et al.\ (2010) have radii $\approx (2 \, {\rm pc}) / 300 = 2 \times 10^{16} \, {\rm cm}$
and surface densities $\approx 2 \times 10^{-4} \, {\rm g} \, {\rm cm}^{-2}$.
This means the clumps should propagate through the isotropic layer with little interaction.   
In fact, the clump densities of $\approx 1 \times 10^{-20} \, {\rm g} \, {\rm cm}^{-3}$
are comparable to the gas density in the molecular cloud.
Furthermore, what Ouellette et al.\ (2010) described were the largest clumps seen in 
numerical simulations and observations, and smaller (possibly denser) clumps are also
possible. 
Clumpy supernova ejecta stand a much better chance of penetrating into the molecular cloud 
than isotropic ejecta.


\section{Methods}

Motivated by these arguments, we carried out a suite of numerical simulations of 
the injection of supernova ejecta into the molecular gas at the periphery 
of an H {\sc ii} region.
All simulations were using the 
FLASH 3.2 multidimensional, adaptive mesh refinement (AMR) code (Fryxell et al.\ 2000) 
that solves the Riemann problem on a Cartesian grid with a directionally-split  
Piecewise Parabolic Method (PPM) (Colella \& Woodward 1984; Colella \& Glaz 1985; 
Fryxell et al.\ 1989).  
We initialized the problem with a planar contact discontinuity at $x=0$, 
separating warm ionized gas in the H {\sc ii} region from
colder neutral gas in the molecular cloud (see Figs. 2 and 3).
Both media are assumed to be uniform in density and temperature, and in pressure
equilibrium at $1.8 \times 10^{-10} \, {\rm dyn} \, {\rm cm}^{-2}$.
The gas in the H {\sc ii} region is assumed to have density
$1.7 \times 10^{-22} \, {\rm g} \, {\rm cm}^{-3}$ and temperature 8000 K, and to 
be fully ionized with mean molecular weight $\mu = 0.6$. 
The molecular gas is assumed to have a  density $3.3 \times 10^{-20} \, {\rm g} \, {\rm cm}^{-3}$
and temperature 40 K. 
For simplicity and to better approximate the equation of state of molecular gas 
when it is shocked, we  set its molecular weight to 0.6 as well.
For the same reasons, we set the ratio of specific heats in both regions to be 
$\gamma = 5/3$.
Thus we do not treat energy losses from dissociation and ionization
in our simulations, but these are probably unimportant as the energies 
are small compared to the kinetic energies of the supernova ejecta material.
 
In the time before the explosion, the massive progenitor will launch a D-type ionization front  which drives a shock that compresses 
gas several $\times 0.1 \, {\rm pc}$ ahead of the  ionization front (Spitzer 1978). 
We do not include this shock in our simulations, deferring a more exact treatment for
future work.
Instead, the two media are assumed to be static before the 
introduction of ejecta. 
Turbulent motions and density inhomogeneities in the cloud are also neglected, as is 
the bulk (outward) motion of the shocked molecular gas.
These motions have magnitudes $\sim 1 \, {\rm km} \, {\rm s}^{-1}$, much smaller than 
the speeds of the ejecta and gas shocked by the ejecta. 
Again, inclusion of these effects is deferred for future work.

Supernova ejecta enter the molecular cloud by moving in the $+x$ direction. 
Isotropic ejecta are modeled as a planar pulse of material on the three-dimensional
Cartesian grid, while clumps of ejecta are modeled as spherical bodies.
We model only a single clump during each simulation. 
Based on the discussion of clumps in the Cas A supernova remnant (\S 2.2), the fiducial 
values for the clumps' masses and radii are $M = 10^{-4} \, M_{\odot}$ and 
$R = 5 \times 10^{15} \, {\rm cm}$, which correspond
to a density $\rho_{\rm ej} = 3.8 \times 10^{-19} \, {\rm g} \, {\rm cm}^{-3}$.
We assume a typical speed of $V = 2000 \, {\rm km} \, {\rm s}^{-1}$.
This is of the same order as the expansion rate of the Cas A supernova remnant
(Chevalier \& Liang 1989) and of many clump velocities, although many clumps also
move faster than this (Fesen et al.\ 2001).
There is considerable uncertainty associated with these values, and we consider a range
of input parameters about these fiducial values.
The temperatures inside the clumps are set to 100 K, due to the fact that they are
dense and cool effectively. 
The exact value is unimportant because each clump will reach very high temperatures
when they are heated by an internal reverse shock after entering the molecular cloud. 

The small clump sizes and large penetration distances we are interested in together
demand a high numerical resolution.
For example, to resolve a clump of size $\simeq 10^{16} \, {\rm cm}$ using 
just 5 zones, on a grid $1 \, {\rm pc}$ across, requires an effective resolution
of $1500^3$.
In our fiducial runs, we choose the refinement level of the adaptive mesh such 
that the smallest resolved scale, $l_{\rm R}$, is $ 2 \times 10^{15}$ cm. 
Our runs were primarily 3D, although we did conduct some high-resolution
2D cylindrically-symmetric runs for comparison, as discussed in \S 4.4.4.
In addition to the hydrodynamic equations, we evolve a scalar field, 
$C({\bs r}, t)$, representing the concentration of heavy elements in an 
ejecta clump.   
Rather than assign to the clump a specific composition from a particular zone
within the supernova, we use this scalar as a generic tracer of ejected material, 
initializing it to be unity inside a clump and zero elsewhere. 
The scalar field is then passively advected in the flow and provides important 
information about the transport and the mixing status of the ejecta material in 
the molecular cloud. 

Cooling of the shocked gas is an important effect and is included in our runs.
The postshock temperatures following the passage of a shock with speed 
$\approx 10^{3} \, {\rm km} \, {\rm s}^{-1}$ are $\gtsimeq 10^{7} \, {\rm K}$,
implying emission of X-ray photons. We assume a cooling function over a 
temperature range $10^{4} - 10^{9} \, {\rm K}$ from the tables compiled by 
Wiersma et al.\ (2009), used in the code CLOUDY (Ferland et al.\ 1998), 
assuming local thermodynamic equilibrium, and solar metallicity. Outside this 
temperature range, we set the cooling rate to zero. The molecular cloud has 
very nearly solar metallicity, while the clump may have a greater metallicity 
and therefore faster cooling rate. 
We assume the emission is optically thin, an assumption we justify below. 

In Fig.~1,  we plot the cooling timescale, $t_{\rm cool}$, as a function of the temperature 
$T$. Here $t_{\rm cool}$ is defined as the time for the gas to radiate away half of 
its internal energy and is calculated from the adopted cooling function $\Lambda(T)$
as $t_{\rm cool} \equiv 3 n k T/ (4 n_{\rm e} n_{\rm H} \Lambda (T) )$,
where $n_{\rm e}$ and  $n_{\rm H}$ are the electron and proton number densities
and $n = \rho/\mu m_{\rm H}$ is the total number density. 
In Fig.~1, the total number density is normalized to a fiducial value of $10^5 \, {\rm cm}^{-3}$ 
because $n = 1.3 \times 10^5$ cm$^{-3}$ for the preshock density of 
$3.3 \times 10^{-20} \, {\rm g} \, {\rm cm}^{-3}$ adopted in our simulations. 
The top $x$-axis gives the speed of a shock that results in a postshock 
temperature corresponding to that shown at the bottom axis. Following the 
passage of a shock with velocity $\approx 2000 \, {\rm km} \, {\rm s}^{-1}$,
cooling times are initially on the order of a few $\times 10^2$ years. The 
cooling timescales drop rapidly with decreasing temperature, though, and become
$< 10 \, {\rm yr}$ below temperatures of $3 \times 10^6 \, {\rm K}$.
The increasingly fast cooling rates at low temperatures potentially can give rise to a 
thermal instability, as discussed below. 

\begin{figure}[ht]
\centerline{ 
\includegraphics[width=1.0\columnwidth]{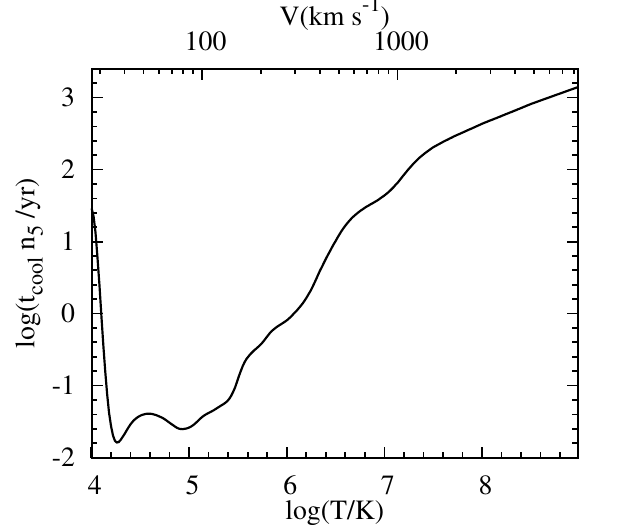} 
} 
\caption[] 
{Cooling timescale as a function of temperature. The upper axis shows 
the shock speed that would give rise to the corresponding temperature.  The 
term $n_5$ is the total number density normalized to $10^5 \, {\rm cm}^{-3}$. }
\label{velocity} 
\end{figure} 

\section{Results} 

\subsection{Isotropic Ejecta}


In our first set of numerical experiments,  we investigate the interaction of the supernova 
ejecta with the molecular cloud assuming an isotropic expansion of the ejecta material.  
We model the isotropic ejecta as a planar pulse, and the initial condition of the ejecta is set to be a uniform slab, moving toward the molecular cloud 
with a velocity of $2000 \, {\rm km} \, {\rm s}^{-1}$. 
The thickness and the density of the slab are set to $1.3 \times 10^{18} \, {\rm cm}$ 
and $6.3 \times 10^{-23} \, {\rm g} \, {\rm cm}^{-3}$, respectively. 
These parameters correspond to a total ejecta mass of 20 $M_{\odot}$ 
distributed uniformly in a spherical shell of thickness  $\simeq 0.4$ pc, 
located at a distance of 2 pc from the explosion center.  
As mentioned earlier, isotropically expanding ejecta will have very low density; 
the ejecta density used here is $\sim 500$ times smaller than the adopted 
density for the molecular cloud. 
We placed this slab in the H {\sc ii} region with its right edge initially located at a 
distance of $1 \times 10^{17} \, {\rm cm}$ from the contact discontinuity at $x=0$.
We applied the outflow boundary condition to all the six sides of the simulation box. 
Once the ejecta encounters the molecular cloud at $x=0$, a contact discontinuity between 
ejecta gas and molecular gas develops, and a shock is driven into the molecular gas.
Although our simulations were three-dimensional and allowed for complex structures,
both the contact discontinuity and the shock remained planar and essentially one-dimensional.
Because of the lack of complex structure we do not present figures depicting these simulations.

In our first simulation, we neglected the radiative cooling.
We find in this case that the high pressure in the postshock gas efficiently decelerates 
the low-density ejecta, and finally pushes all the ejecta material back to the $x<0$ region. 
In the first 200 yr, a small fraction (20\%) of the ejecta material manages to pass the 
$x=0$ plane, but it only reaches a negligibly small distance ($x < 5 \times 10^{16}$ cm)
before it starts to move backward at $t= 200$ yr, due to the large postshock pressure. 
By 500 yr, all of the ejecta material has ``bounced" back into $x<0$. 
In these simulations without cooling, the ejecta does not remain in contact with the molecular
gas long enough to mix into it, even if the structure had not remained planar.

We next conducted a second simulation identical to the first but including radiative 
cooling.
As pointed out earlier, the cooling timescale  in the postshock gas is quite short, 
$\simeq 100$ yr for our fiducial parameters, so it is unsurprising that cooling changes the 
dynamical behavior of the gas.
The radiative cooling is found to reduce the postshock pressure, which, in a cooling timescale, 
becomes significantly smaller than in the case neglecting cooling. 
Our simulation shows that the efficient cooling gives rise to the formation of a dense shell behind 
the shock in a few hundred years. 
The molecular gas swept up by the shock piles up in the shell, whose width increases with time. 
Due to the reduction of the postshock pressure by cooling, the ejecta material can follow the shell, 
and continuously fill up the space behind the condensed gas. 
In other words, when the radiative cooling is accounted for, the molecular gas can be compressed
and pushed by the ejecta.  
Despite this, both the contact discontinuity and the shock are found to remain planar, with 
no mixing of ejecta into the shocked molecular gas.
We observed the ejecta concentration field, $C(\bs{r}, t)$, to remain at its initial value 
($= 1$) on the left side of the discontinuity and to be zero on the right side. 
As may be expected, the contact discontinuity is found to move with a speed 3/4 of the shock
velocity at all times.
The contact discontinuity persisted throughout the run and no shear or shear-related instabilities
were observed to arise.
The RT instability was not observed either, nor should it arise, because the 
density of the ejecta is significantly lower than the molecular gas. 

In these runs we observed no instabilities at the contact discontinuity between 
ejecta gas and the molecular cloud, meaning there is no mechanism to mix the 
ejecta gas with the molecular cloud material that will go on to form protostars.
One effect that may alter this picture is pre-existing turbulence in the molecular 
cloud, wherein random motions may perturb the shock front. 
A more likely mechanism is pre-existing density inhomogeneities in the molecular 
cloud such as cloud cores. The interaction of a cloud core with a sweeping shock 
has been explored by Boss and collaborators (e.g., Boss et al.\ 2008, 2010, Boss and Keiser 2010). 
Shear and Kelvin-Helmholtz instabilities would arise as the shock sweeps past 
these cloud cores. Future work will further explore these alternative scenarios. In the present work,
we find that if ejecta are homogeneously distributed, then 
mixing of ejecta material into the molecular cloud gas is unlikely, because of the low
density of the ejecta and the large density contrast it has with the molecular gas.

\subsection{Clumpy Ejecta with Cooling Neglected} 

\begin{figure*}
\centerline{
\includegraphics[width=2.0\columnwidth]{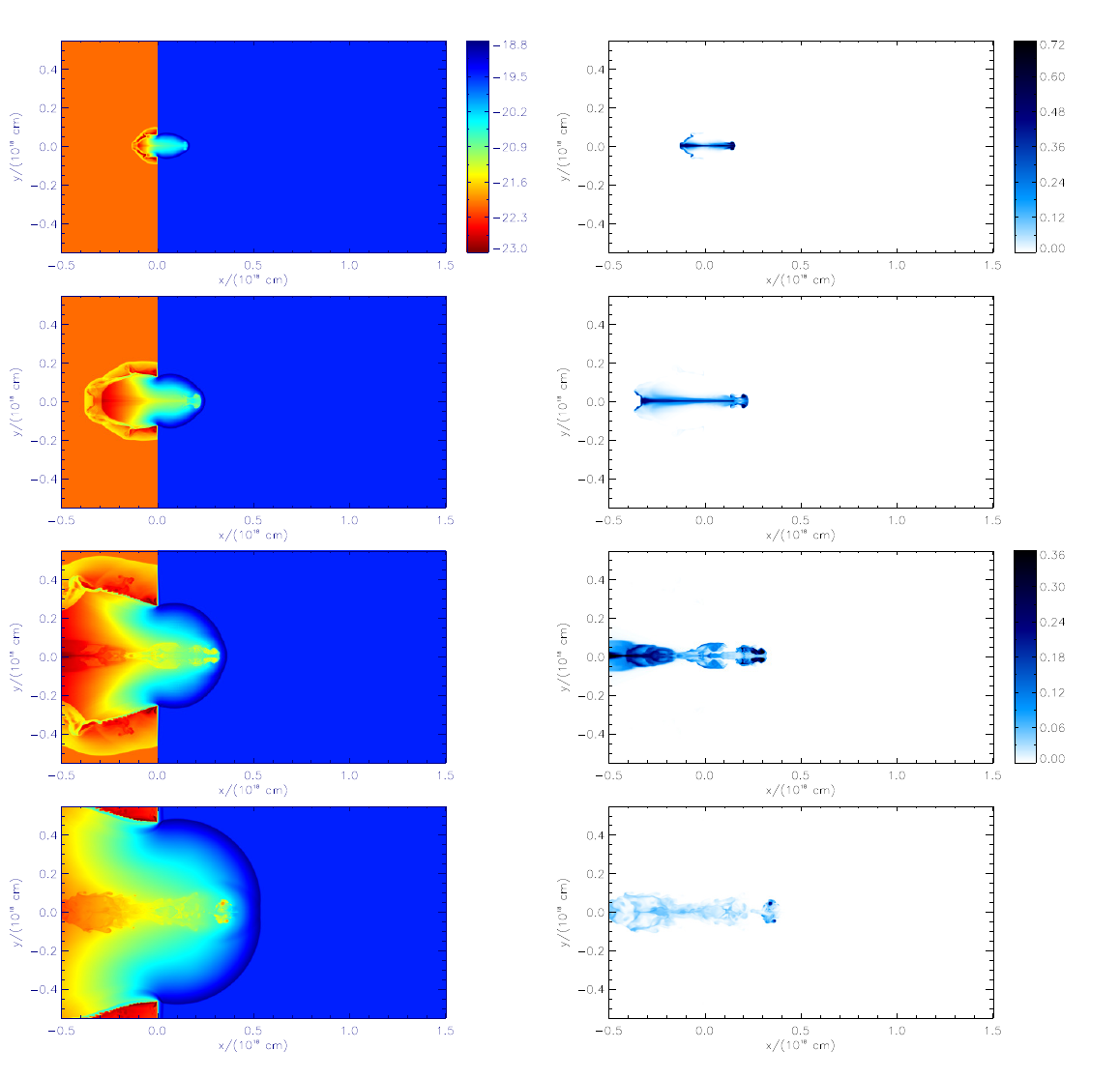}
}
\caption[]{
Evolution of the gas density (left panels) and the ejecta concentration
(right panels), for the case in which radiative cooling is neglected.
Density contours are shown using colors corresponding to the numbers 
on the scale bar, which are the logarithms of the density, expressed in units of 
${\rm g} \, {\rm cm}^{-3}$.  Concentration contours are shown using contours ranging 
from zero ejecta fraction (white) to an ejecta fraction near unity (dark blue).
From top to bottom, the four times depicted are $t = 100$, 500, 2400 and 10,000 years
following the impact of the ejecta clump with the molecular gas.  For clarity, the
bottom two panels use a different color range for the ejecta concentration field.
Parameters are described in the text.} 
\label{nocooling}
\end{figure*}    

The next simulation we present considers clumpy ejecta, but neglects radiative 
cooling in the post-shock regions.The three-dimensional Cartesian 
computational domain is a cubic box of size $3 \times 10^{18} \, {\rm cm}$ on a side, with a base grid of $48^3$ cells, 
$6 \times 10^{16}$ cm on a side.  Again the outflow boundary condition is chosen for each 
side of the grid. Using the standard density refinement criteria 
available with FLASH, we allow for 5 additional levels of refinement, so that the smallest resolved scale is $2 \times 10^{15} \, {\rm cm}$,
corresponding to an effective resolution of $\simeq 1500^3$. 
A spherical clump of ejecta, with radius $R = 5 \times 10^{15} \, {\rm cm}$ and
density $\rho_{\rm ej} = 3.8 \times 10^{-19}$ g cm$^{-3}$, is initially located at 
${\bs r} = (-5\times 10^{16}, 0, 0)$ (i.e., slightly to the left of the contact 
discontinuity at the $x=0$ plane).  
It moves toward the molecular cloud with an initial velocity 
$V = 2000 \, {\rm km} \, {\rm s}^{-1}$.

The results of the simulation are presented in Fig.~\ref{nocooling}.
In this figure, only part of the full computational domain is shown.
The left four panels plot the evolution of the density field on a logarithmic 
scale on the $x$-$y$ plane at four different times (100 yr, 500 yr, 2400 yr and
10,000 yr). The clump is seen to drive a strong shock as it enters the molecular 
cloud.   The shock front appears to  be an ellipsoid elongated in $x$-direction, 
and later evolves toward  a nearly spherical shape due to lateral expansion
driven by the high pressure in the post-shock regions. 
With time, the shock front sweeps up more mass and moves deeper into the molecular cloud. 

Due to the asymmetry of the shock front, shear flow emerges behind the shock, 
which gives rise to the Kelvin-Helmholtz (KH) instability. 
The KH vortices disperse and help mix the ejecta in the postshock region. 
The right panels in Fig.~\ref{nocooling} show the concentration field, $C$, of the ejecta 
material at times $t=100$, 500, 2,400 and 10,000 yr, respectively. 
Here we clearly see  that KH vortices stretch the ejecta and spread them laterally. 
On the other hand, the RT instability was not observed around the ejecta.
This may be due to the fact that we could only achieve a numerical resolution
corresponding to 5 zones across the clump.
Mac Low \& Zahnle (1994), in their 2D cylindrical simulations of the breakup of
comet Shoemaker-Levy 9 in Jupiter's atmosphere, have argued that 25 zones across
the projectile are required to resolve RT instabilities.
We return to this issue in \S 4.5.

The right panels of Fig.~\ref{nocooling} illustrate that only some of the ejecta
closely follow behind the shock. 
A significant fraction of the ejecta lags behind the shock, and some  fraction of 
this material even moves backward and starts to leave the computational domain
by $t \simeq 1000$ yr.  
The high density and temperature behind the shock give rise to a strong negative pressure 
gradient in the $x$-direction, which drives the backward motion of the ejecta.
At $t=2400$ yr, only half of the ejecta remains within the molecular cloud 
(i.e., has  $x \geq 0$), and 20\% has already been lost from the computational domain.
The fraction of the ejecta mass remaining within the molecular cloud decreases with time.
By $t = 10^4$ yr, only 25\% of the original ejecta clump mass remains on the computational
domain, and only 15\% remains in the molecular cloud (Fig.~\ref{nocooling}).
By $t = 3 \times 10^{4}$ yr, only 2\% of the original clump mass remains in the cloud.
We expect that even less of the clump mass would be injected into the cloud at longer
times. 

We also found the same general trend for the clump mass to ``bounce" out of the cloud 
in 2D simulations of the same problem using cylindrical coordinates, as already seen
in the isotropic ejecta case. 
These effects are not sensitive at all to the clump parameters or the numerical 
resolution.
We conclude that in the case where radiative cooling is neglected, clumps of ejecta 
are not injected into the molecular cloud.
Instead, similar to an inelastic collision between a ball and fixed wall,
they are effectively bounced off of the molecular cloud surface by the 
large postshock pressure.

\subsection{Clumpy Ejecta with Cooling Included}

\begin{figure*}
\centerline{
\includegraphics[width=2.\columnwidth]{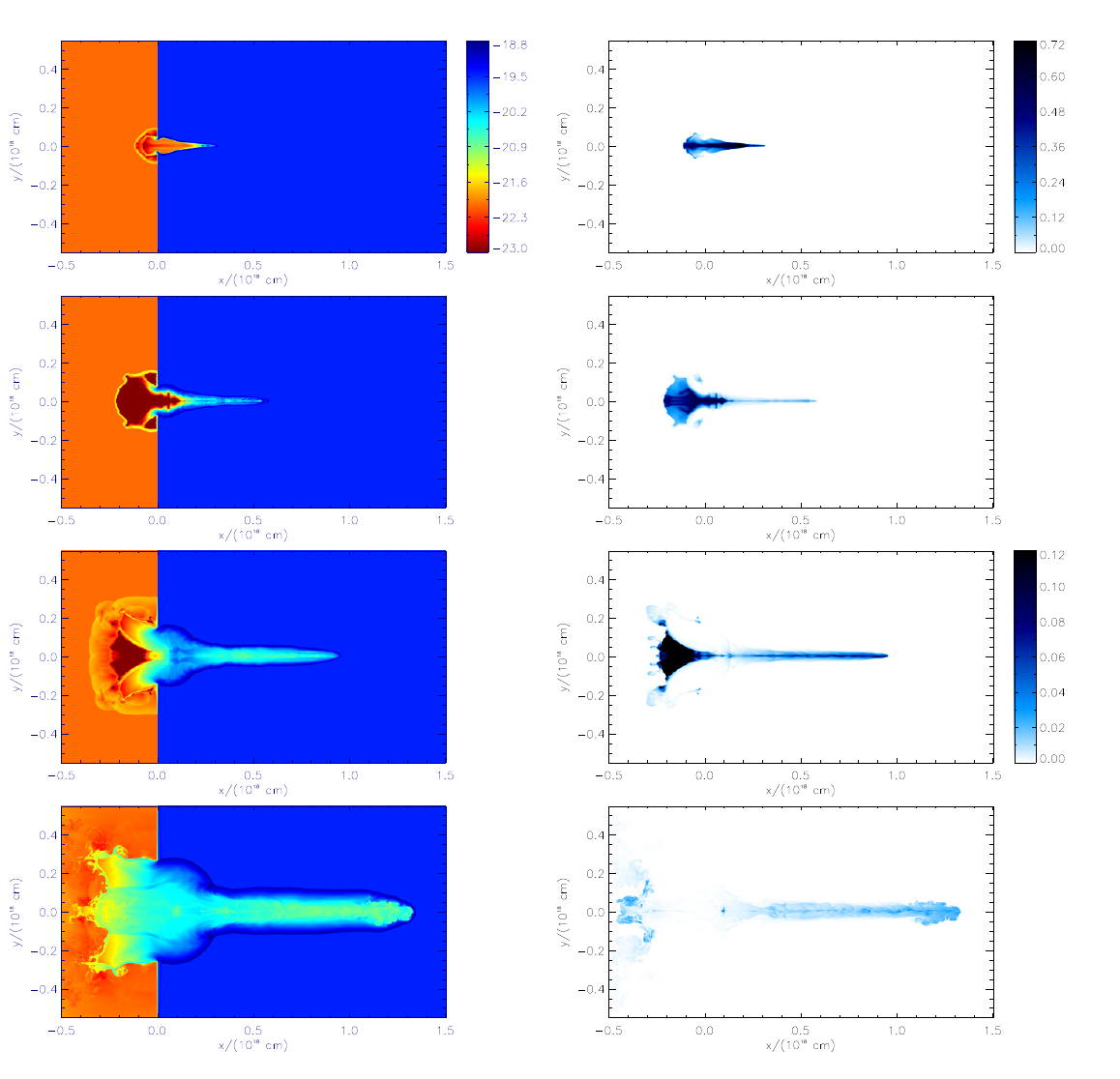}
}
\caption[]{
Evolution of the gas density (left panels) and the ejecta concentration
(right panels), for the case in which radiative cooling is included.
Density contours are shown using colors corresponding to the numbers 
on the scale bar, which are the logarithms of the density, expressed in units of 
${\rm g} \, {\rm cm}^{-3}$.  Concentration contours are shown using contours ranging 
from zero ejecta fraction (white) to an ejecta fraction near unity (dark blue).
From top to bottom, the four times depicted are $t = 100$, 500, 2400 and 10,000 years
following the impact of the ejecta clump with the molecular gas.  For clarity, the
bottom two panels use a different color range for the ejecta concentration field.
}
\label{cooling}
\end{figure*}

In the preceding simulations not including cooling, clumpy ejecta manage to move farther
(to larger $x$) than the isotropic ejecta, but ultimately they are expelled from the molecular 
cloud by the high post-shock pressures without mixing into the cloud.
We expect that if cooling is significant, it should reduce this pressure and allow deeper 
penetration of the clumpy ejecta into the cloud.
Including cooling as outlined in \S 3, we
find that the high density of the molecular cloud does result in a significant drop
in postshock temperature and pressure, because of the very short cooling timescale.
For our adopted cooling rates, molecular gas with a preshock number density
$n_{\rm H2} = 10^4 \, {\rm cm}^{-3}$, shocked by a $2000 \, {\rm km} \, {\rm s}^{-1}$
shock to a temperature $\approx 5 \times 10^{7} \, {\rm K}$ will cool in only
$\approx 100 \, {\rm yr}$.
This is comparable to the dynamical timescales in the problem, which we estimate
as $(10^{18} \, {\rm cm}) / (2000 \, {\rm km} \, {\rm s}^{-1})$ $\sim 100 \, {\rm yr}$
(the time for a clump to move a significant distance through the cloud).
Our simulations confirm that the ejecta do indeed reach significant depths in
the cloud when radiative cooling is included. 

The left panels of Fig.~\ref{cooling} show the density field at four snapshots in time, at 
$t=100$, 500, 2400, and 10,000 yr.
All parameters are the same as in the fiducial simulation outlined in \S 4.2, 
except that radiative cooling is now included.
Our base resolution was $6 \times 10^{16} \, {\rm cm}$ and our effective resolution
was $2 \times 10^{15} \, {\rm cm}$, which is 5 times smaller than the clump diameter.   
To allow for higher penetration depths as well as reduce the computational cost per
runs  we extended the computational domain to $4.5 \times 10^{18} \, {\rm cm}$ in the $x$ 
direction and reduced it to $2 \times 10^{18} \, {\rm cm}$ in the $y$ and $z$ directions.
Fig.~\ref{3dplot}a represents a 3D rendering of the same output, at a time 
$t = 2400$ years. 

We find that an early phase exists where the dynamical behavior is very similar 
to the case with no cooling. 
The phase lasts for about a cooling timescale, or $\ltsimeq 50 \, {\rm yr}$.
During this initial phase, the ejecta clump produces a shock front in the molecular
gas with an ellipsoidal shape, which then evolves to a nearly spherical shape on the 
left (trailing) side of the density field at later times, as seen in the left bottom 
two panels of Fig.~\ref{cooling}.
The post-shock pressure converts part of the kinetic energy in the $x$-direction 
into lateral expansion. 
We find that this early evolution phase, lasting 1 cooling timescale, plays a 
crucial role in determining how far the ejecta is delivered into the molecular 
cloud (see \S 4.4).  

At times greater than the cooling time, the geometry of the flow changes significantly.
Radiative cooling is now significant and the postshock 
pressure is reduced, allowing the ejecta to more closely follow the shock front.
As seen in the top panel of Fig.~\ref{cooling}, the ejecta opens up a narrow channel into the 
molecular cloud at $t \simeq 100 \, {\rm yr}$.
At about this time, the shock front propagation in the $x$-direction becomes driven 
essentially by the momentum of the ejecta alone, and the ejecta motion appears 
to be ballistic. 
Due to the lack of strong lateral pressure gradients, the expansion in the lateral  
direction is weak. 
Initially, both the length and width of the channel increase with time (akin to a 
Mach cone).
After $\approx 5000 \, {\rm yr}$, the ejecta have lost significant momentum and the 
shock does not move significantly in the $x$-direction, even as the channel width 
keeps growing in the $y$ and $z$-directions. 
The shock front reaches a distance of $\approx 0.6 \, {\rm pc}$
in the molecular cloud at 30,000 yr. 

The spatial distribution of the supernova material in the postshock region is shown in 
the right panels of Fig.~\ref{cooling}.
The ejecta are seen to follow the shock front more closely than in the case without
cooling.
The supernova material is much more evenly distributed along the channel, and only
a small fraction is expelled from the molecular cloud.

Note that the right column of Fig.~\ref{cooling} plots the concentration of supernova material 
in the gas.
Although the central two panels leave the impression that a considerable amount of ejecta 
is bounced out of the cloud to $x \le 0$, the dark blue regions on the left with high
ejecta concentration actually have little ejecta mass, because the gas density is 
extremely low in this region. 
In fact, during the entire simulation (up to 30,000 yr), nearly all (93\%) of the 
ejecta remain in the computational domain, and a very high fraction (86\%) 
are delivered deep into the molecular cloud.
Furthermore, the delivery depth appears to be controlled by the 
$x$-momentum per unit area of the clump, a point we return to in \S 4.4 below. 

\begin{figure}[ht]
\centerline{
\includegraphics[width=1.0\columnwidth]{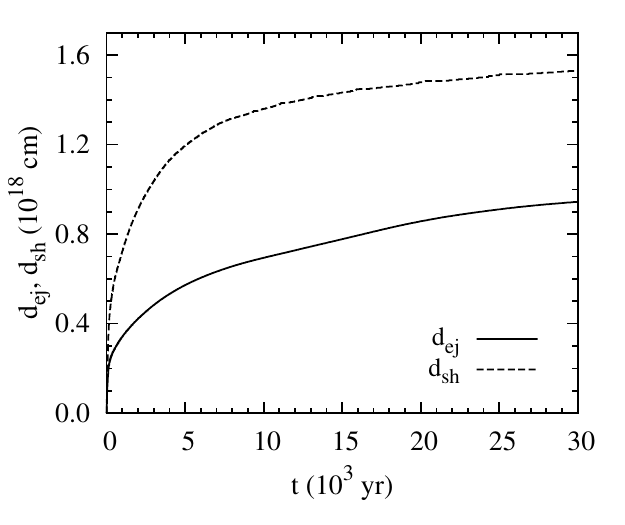}
}
\caption[]{The ejecta delivery distance, $d_{\rm ej}$, and the position, $d_{\rm sh}$, 
of the leading shock front as a function of time. The clump velocity, density 
and size are, respectively, 2000 km s$^{-1}$, $3.8 \times 10^{-19}$  g cm$^{-3}$, 
and  $0.5 \times 10^{16}$ cm.} 
\label{delivery}
\end{figure}

To more precisely quantify the delivery of ejecta, we measure the mass center of the 
ejecta material in the $x$-direction, 
$d_{\rm ej} \equiv M^{-1} \, \int \rho({\bs r}, t) C({\bs r}, t) \, x \, d{\bs r}$, 
where $\rho({\bs r}, t)$ is the gas density, $C({\bs r}, t)$ the ejecta concentration, 
and $M=\int \rho({\bs r}, t) C({\bs r}, t) d{\bs r}$ is the total mass of the clump material.  
In Fig.~\ref{delivery}, we plot $d_{\rm ej}$ as a function of time in our fiducial run with
cooling. 
Also shown is the distance, $d_{\rm sh}$, that the leading shock front has travelled into 
the cloud in the $x$-direction. 
The shock appears to stall after about 15,000 yr, at a distance
$d_{\rm sh} \approx 0.5 \, {\rm pc}$. 
The ejecta themselves are delivered more or less evenly to distances up to $d_{\rm sh}$ ,
with an average depth of $d_{\rm ej} \approx 0.3 \, {\rm pc}$.

As in the case without cooling, KH instabilities are seen in the channel at late 
times, which help mix ejecta with the molecular gas.
At $t = 10,000 \, {\rm yr}$, this mixing has significantly reduced the maximum concentration 
of the ejecta, as seen from the color scale used in the bottom right panel of Fig.~\ref{cooling}.
As before, RT instabilities were not observed around the ejecta material in the channel. 
It is possible that the RT instability may have been suppressed by the 
efficient radiative cooling, which reduces the pressure in the postshock regions and 
thus leads to a slower relative acceleration between the ejecta and the shocked gas. 
It is also likely that emergence of the RT instability requires higher numerical resolution 
than we achieved, since RT instabilities were not observed in the case without
cooling, either.
We do observe an RT-like (Richtmyer-Meshkov) instability as the dilute high-pressure gas 
is pushed back to the warm ionized gas on the left side (see the third right panel in 
Fig.~\ref{cooling}).
We also observe what may be a cooling instability, during the early phase when the thermally 
unstable gas (with temperatures in the range $10^4 - 10^7$ K) emerge into the postshock 
region (see discussion in \S 4.4.3).  

We have assumed the emission of radiation is in the optically thin limit.
Our simulations allow us to test this assumption post priori.
Along the lateral direction, the column density of shocked gas is dominated by the 
dense gas at the boundary of the channel.
The width of the channel itself is $< 10^{17} \, {\rm cm}$ and the density within
it is $< 10^{-20} \, {\rm g} \, {\rm cm}^{-3}$, making the column density of gas
across the column $< 10^{-3} \, {\rm g} \, {\rm cm}^{-2}$.  
The shell of gas at the edge of the channel has a width $< 2 \times 10^{16} \, {\rm cm}$
and a density $< 10^{-19} \, {\rm g} \, {\rm cm}^{-3}$, for a column density
$< 2 \times 10^{-3} \, {\rm g} \, {\rm cm}^{-2}$.
At the X-ray wavelengths that dominate the cooling, scattering is due to free electrons
and absorption is due to the inner shell transitions of heavy atoms.
The column density required for an optical depth of 1 due to Thomson scattering
is $\sim 3 \, {\rm g} \, {\rm cm}^{-2}$, meaning there is insignificant scattering
of X rays even as they propagate across the channel.
Including absorption (with solar metallicity), X rays with energy 5 keV 
(corresponding to $10^{7} \, {\rm K}$ gas) are not significantly attenuated by 
column densities $\ltsimeq 0.2 \, {\rm g} \, {\rm cm}^{-2}$ (Igea \& Glasgold 1999).
The radiative cooling by emission of X rays in the direction across the channel is 
therefore optically thin.
Along the channel, in the direction toward the H {\sc ii} region, the column density 
is only a factor of 10 larger, $\sim 10^{-2} \, {\rm g} \, {\rm cm}^{-2}$, and also
allows optically thin emission.

In summary, we find that the molecular cloud is dense enough to cool effectively, but is
not so dense as to become optically thick to its own emission.  
Radiative cooling is a significant effect and plays a crucial role in the delivery of 
supernova material into a nearby molecular cloud. 
Radiative cooling reduces the pressure in the postshock regions, meaning the ejecta 
clump is limited only by its forward momentum.  
As it moves ballistically through the cloud it opens a narrow channel. 
For the parameters we consider typical of supernova clumps and the surrounding molecular
cloud, the ejecta moves on average a distance $\sim 0.3 \, {\rm pc}$ into the molecular gas.


\subsection{Parameter Study} 

To test the sensitivity of ejecta delivery on the clump parameters, we conducted a suite 
of simulations with varying initial conditions. 
Our results are summarized in Table 1, which lists the farthest advance of the shock 
front, $d_{\rm sh}$, the centroid of the ejecta, $d_{\rm ej}$, and the fraction of ejecta remaining 
in the molecular cloud ($x > 0$), $f_{\rm inj}$, after 30,000 years, as a function of clump 
velocity $V$, clump density $\rho_{\rm ej}$, and clump radius $R$.
The results are discussed in \S\S 4.4.1-4.4.3 below, which describe the effect of 
varying clump velocity, density, and size from their fiducial values of $V = 2000 \, {\rm km} \, {\rm s}^{-1}$, 
$\rho_{\rm ej} = 3.8 \times 10^{-19} \, {\rm g} \, {\rm cm}^{-3}$, 
and $R = 5 \times 10^{15} \, {\rm cm}$. Radiative cooling was included in each of these simulations.
To assess the numerical convergence, we also conducted a run with an additional refinement level,
yielding an effective resolution of $l_{\rm R} = 1 \times 10^{15}$ cm.  At this resolution, 
the computation is already very expensive. We carried out the simulation with 1024 processors, 
and it lasted for about 8 days, meaning a cost of 200,000 CPU hours for a single run. 
Higher numerical resolutions are desirable, but would be prohibitively expensive for a parameter study.  
We return to the issue of numerical convergence in \S 4.5.

\begin{table*}
\begin{center}
\caption{Parameters studies in 3D simulations of ejecta delivery into molecular cloud}
\label{tbl-1}
\scalebox{1}{
\begin{tabular}{c@{\hspace{3mm}}r@{\hspace{5mm}}r@{\hspace{5mm}}c@{\hspace{7mm}}c@{\hspace{7mm}}r@{\hspace{7mm}}c@{\hspace{7mm}} c}
\hline
\hline
Case & $V$ (km/s) & $\rho_{\rm ej}$ (g cm$^{-3}$) & $R$ (cm) & $M$ ($M_\odot$) &  $d_{\rm sh}$\tablenotemark{1} (cm)  
&  $d_{\rm ej}$\tablenotemark{2} (cm) & $f_{\rm inj}$\tablenotemark{3} \\ 
\hline
 1 &  500 & $3.8 \times 10^{-19}$ &  $ 5\times 10^{15}$  & $1.0 \times 10^{-4}$ &  $9.2 \times 10^{17}$  &  $6.1 \times 10^{17}$ & 98.5\% \\
 2 & 1000 & $3.8 \times 10^{-19}$ &  $ 5\times 10^{15}$  & $1.0 \times 10^{-4}$ &  $1.3 \times 10^{18}$  &  $8.5 \times 10^{17}$ & 94.6\% \\
 3 & 2000 & $3.8 \times 10^{-19}$ &  $ 5\times 10^{15}$  & $1.0 \times 10^{-4}$ &  $1.5 \times 10^{18}$  &  $9.4 \times 10^{17}$ & 86.1\% \\
 4 & 2500 & $3.8 \times 10^{-19}$ &  $ 5\times 10^{15}$  & $1.0 \times 10^{-4}$ &  $7.2 \times 10^{17}$  &  $4.1 \times 10^{17}$ & 83.9\% \\
 5\tablenotemark{4} 
   & 3000 & $3.8 \times 10^{-19}$ &  $ 5\times 10^{15}$  & $1.0 \times 10^{-4}$ &  $\approx 5 \times 10^{17}$ & $\approx 1.7 \times 10^{17}$ & $< 77\%$ \\
 6\tablenotemark{4} 
   & 4000 & $3.8 \times 10^{-19}$ &  $ 5\times 10^{15}$  & $1.0 \times 10^{-4}$ &  $\approx 5 \times 10^{17}$ & $\approx 1.3 \times 10^{17}$ & $< 65\%$ \\
\hline
 7 & 2000 & $1.9 \times 10^{-19}$ &  $ 5\times 10^{15}$  & $0.5 \times 10^{-4}$ &  $4.7 \times 10^{17}$  &  $1.3 \times 10^{17}$ & 67.8\% \\
 3 & 2000 & $3.8 \times 10^{-19}$ &  $ 5\times 10^{15}$  & $1.0 \times 10^{-4}$ &  $1.5 \times 10^{18}$  &  $9.4 \times 10^{17}$ & 86.1\% \\
 8 & 2000 & $7.6 \times 10^{-19}$ &  $ 5\times 10^{15}$  & $2.0 \times 10^{-4}$ &  $2.3 \times 10^{18}$  &  $1.5 \times 10^{18}$ & 93.5\% \\
\hline
 3 & 2000 & $3.8 \times 10^{-19}$ &  $ 5\times 10^{15}$  & $1.0 \times 10^{-4}$ &  $1.5 \times 10^{18}$  &  $9.4 \times 10^{17}$ & 86.1\% \\
 9 & 2000 & $3.8 \times 10^{-19}$ &  $ 7\times 10^{15}$  & $2.7 \times 10^{-4}$ &  $9.3 \times 10^{17}$  &  $5.1 \times 10^{17}$ & 93.4\% \\
10 & 2000 & $3.8 \times 10^{-19}$ &  $10\times 10^{15}$  & $8.0 \times 10^{-4}$ &  $2.5 \times 10^{18}$  &  $1.0 \times 10^{18}$ & 96.2\% \\
\hline
11 & 1000 & $1.9 \times 10^{-19}$ &  $ 5\times 10^{15}$  & $0.5 \times 10^{-4}$ &  $8.9 \times 10^{17}$  &  $5.0 \times 10^{17}$ & 86.8\% \\
 2 & 1000 & $3.8 \times 10^{-19}$ &  $ 5\times 10^{15}$  & $1.0 \times 10^{-4}$ &  $1.3 \times 10^{18}$  &  $8.5 \times 10^{17}$ & 94.6\% \\
12 & 1000 & $7.6 \times 10^{-19}$ &  $ 5\times 10^{15}$  & $2.0 \times 10^{-4}$ &  $1.8 \times 10^{18}$  &  $1.3 \times 10^{18}$ & 97.5\% \\
\hline
 2 & 1000 & $3.8 \times 10^{-19}$ &  $ 5\times 10^{15}$  & $1.0 \times 10^{-4}$ &  $1.3 \times 10^{18}$  &  $8.5 \times 10^{17}$ & 94.6\% \\
13 & 1000 & $3.8 \times 10^{-19}$ &  $ 7\times 10^{15}$  & $2.7 \times 10^{-4}$ &  $2.0 \times 10^{18}$  &  $1.4 \times 10^{18}$ & 97.2\% \\
14 & 1000 & $3.8 \times 10^{-19}$ &  $10\times 10^{15}$  & $8.0 \times 10^{-4}$ &  $2.8 \times 10^{18}$  &  $2.2 \times 10^{18}$ & 98.5\% \\
\hline
15\tablenotemark{5} & 2000 & $3.8 \times 10^{-19}$ &  $ 5\times 10^{15}$  & $1.0 \times 10^{-4}$ &  $5.5 \times 10^{17}$  &  $< 1\times 10^{17}$ & 53.7\% \\
                 3 & 2000 & $3.8 \times 10^{-19}$ &  $ 5\times 10^{15}$  & $1.0 \times 10^{-4}$ &  $1.5 \times 10^{18}$  &  $9.4 \times 10^{17}$ & 86.1\% \\
16\tablenotemark{5} & 2000 & $3.8 \times 10^{-19}$ &  $ 5\times 10^{15}$  & $1.0 \times 10^{-4}$ &  $2.1 \times 10^{18}$  &  $1.3 \times 10^{18}$ & 96.6\% \\
\hline
\\
\multicolumn{8}{l}{$^1$Distance to which leading shock front advances.}\\
\multicolumn{8}{l}{$^2$Mean depth into the cloud to which ejecta penetrates.}\\
\multicolumn{8}{l}{$^3$Fraction of ejecta remaining in cloud at 30,000 years.}\\
\multicolumn{8}{l}{$^4$Cases 5 and 6 were stopped at 6000 years.  Results are extrapolations based on their behavior to that point; see text.}\\
\multicolumn{8}{l}{$^5$All cases used numerical resolution $l_{\rm R} = 2 \times 10^{15} \, {\rm cm}$, except cases 15 ($l_{\rm R} = 4 \times 10^{15} \, {\rm cm}$)
and 16 ($l_{\rm R} = 1 \times 10^{15} \, {\rm cm}$).}\\
\end{tabular}
\vspace{-0.15in}
}

\end{center}
\end{table*}

\subsubsection{Effect of Clump Velocity}

We first study the dependence of the ejecta delivery on the initial clump 
velocity, $V$. 
The clumps observed in the Cas A supernova remnant span a wide range of velocities. 
Most of the clumps are observed as they pass through the reverse shock at a distance
$\sim 2'$ (2 pc at 3.4 kpc); if the explosion occurred in 1680, their expansion velocity 
would be $6000 \, {\rm km} \, {\rm s}^{-1}$. 
The clumps observed by Fesen et al.\ (2001) show emission from oxygen, nitrogen and sulfur,
possibly suggestive of arising from the intermediate layers of the supernova. 
Clumps from deeper in the explosion would not be moving as rapidly and would not have 
reached the reverse shock yet, and so would be unshocked and invisible.
The numerical simulations of Kifonidis et al.\ (2006) suggest that the Fe-rich material 
from deep inside the progenitor explodes outward at $\approx 3300 \, {\rm km} \, {\rm s}^{-1}$.
These facts suggest that a relevant range of velocities to explore would extend 
 up to $6000 \, {\rm km} \, {\rm s}^{-1}$. 

We carried out four 30,000-year runs with velocities in the range 
$500 \, {\rm km} \, {\rm s}^{-1}$ to $2500 \, {\rm km} \, {\rm s}^{-1}$.  
We also carried out  two cases with higher velocity, $V = 3000 \, {\rm km} \, {\rm s}^{-1}$
and $V = 4000 \, {\rm km} \, {\rm s}^{-1}$.
Due to their shorter Courant times, these runs were more computationally expensive
and were terminated at $t = 6000$ years.
These results were then extrapolated to infer the behavior at even higher velocities.

In Fig.~\ref{velocity}, we show $d_{\rm ej}$ as a function of time for runs with 
$V = 500$, 1000, 2000 and 2500 ${\rm km} \, {\rm s}^{-1}$.
The delivery distance appears to be greatest for $V \approx 2000 \, {\rm km} \, {\rm s}^{-1}$,
and to decrease at lower or higher velocities. 
At lower clump velocities, the postshock temperatures are lower, and the radiative 
cooling is faster. 
For example, for a clump with $V=500 \, {\rm km} \, {\rm s}^{-1}$, the cooling timescale in 
the postshock region is $\ltsimeq 10$ years. 
For these low-velocity cases, the ejecta motion of the clump appears to be ballistic at 
all times, indicating that the momentum-driven phase starts almost immediately after the 
clump enters the molecular cloud.  
At no stage is there a spherical component to the shock near the edge of the cloud,
in contrast to the higher-velocities cases.
Based on this, conservation of momentum would suggest that at low velocities the delivery 
distance scales with the clump velocity, a trend that is observed in Table 1.

\begin{figure}[ht]
\centerline{
\includegraphics[width=1\columnwidth]{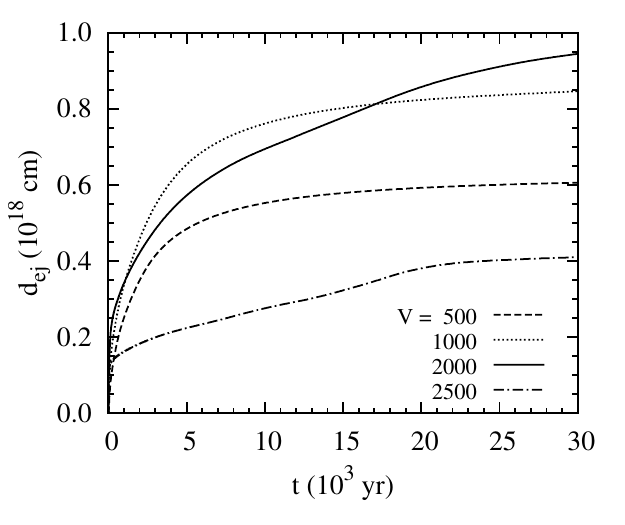}
}
\caption[]
{Variation of $d_{\rm ej}$ with time, for the four different clump velocities
$V=500$, 1000, 2000 and 2500 ${\rm km} \, {\rm s}^{-1}$. 
The clump density was set to $3.8 \times 10^{-19} \, {\rm g} \, {\rm cm}^{-3}$, 
and the clump radius was set to $0.5 \times 10^{16} \, {\rm cm}$.}
\label{velocity}
\end{figure}  

At clump velocities $\gtsimeq 2000 \, {\rm km} \, {\rm s}^{-1}$,    
this trend of increasing penetration distance with increasing speed is counteracted 
by the fact that radiative cooling becomes less effective with increasing 
velocity and postshock temperature.
The postshock pressure remains high for about 1 cooling time, which increases
in duration with increasing clump velocity.
The postshock pressure has the effect of decelerating the ejecta in the 
$x$ direction and laterally spreading its mass and $x$-momentum across a larger
cross-sectional area.
We have computed the root-mean-square displacement of the ejecta material in the
lateral direction and confirm that with increasing clump velocity
the $x$-momentum becomes less concentrated along the penetration 
axis at $y, z=0$.
Because of this lateral spreading, the clump sweeps up more mass
and decelerates more quickly, only becoming ballistic after 1 cooling time.
The shape of the shock front also changes in time differently as the clump 
velocity increases.
For $V = 2500 \, {\rm km} \, {\rm s}^{-1}$, a narrow channel does not form,
and the shock front remains in the ellipsoidal shape for a longer time.  
Because larger clump velocities are associated with longer cooling times,
the general trend is for the penetration depth to decrease with increasing
velocity.
Ejecta with initial velocity $2000 \, {\rm km} \, {\rm s}^{-1}$ penetrate
to $9.5 \times 10^{17} \, {\rm cm}$ by 30,000 years, but clumps with velocity
$2500 \, {\rm km} \, {\rm s}^{-1}$ penetrate only to $4.0 \times 10^{17} \, {\rm cm}$
in the same time (Fig.~\ref{velocity}).
For $V = 2500 \, {\rm km} \, {\rm s}^{-1}$, only 83.9\% of the ejecta remain
in the cloud at 30,000 years.
Extrapolating the behavior of runs 5 and 6 at 6000 years forward in time to 
30,000 years, we estimate that the fraction of ejecta in the cloud at those times 
would be even lower, although a majority of the ejecta would still remain in the
cloud. 
At higher velocities $V \approx 6000 \, {\rm km} \, {\rm s}^{-1}$, we anticipate
that the supernova material would be injected to a shallower depth 
($\ltsimeq 1 \times 10^{17} \, {\rm cm}$), and the fraction remaining in the 
cloud would be reduced ($\ltsimeq 50\%$),  for the fiducial clump size and 
density.

In summary, we find that the clump velocity plays a critical role in setting
how far the ejecta penetrate, and what fraction of the clump mass remains 
in the molecular cloud. 
For our fiducial parameters, the distance to which the average supernova material
is injected is greatest, $d_{\rm ej} \approx 0.3 \, {\rm pc}$, for clump velocities
in the range $V \approx 1000 - 2000 \, {\rm km} \, {\rm s}^{-1}$.
At lower velocities, $d_{\rm ej}$ is set by the clump's momentum, and increases
with $V$.   
At higher velocities, the postshock temperatures and the cooling time increase
with $V$, and the clump is spread out laterally and decelerated more rapidly.
As a general rule, though, the fraction of ejecta that remains in the molecular
cloud ($x > 0$) is greatest at low velocities and decreases with increasing $V$.
At low velocities ($V \leq 500 \, {\rm km} \, {\rm s}^{-1}$), the ejecta do not 
penetrate very far but because the cooling timescales are so short, this material 
remains embedded in the molecular gas.
At high velocities ($V \geq 2500 \, {\rm km} \, {\rm s}^{-1}$), the combination
of low $d_{\rm ej}$ and the long cooling timescales and the geometry of the 
shock allow supernova material to escape to the H {\sc ii} region more easily.
Note however that these results are  only for the fiducial choices of size and density.
The most important insight from the above calculations is that the injection
efficiency is tied to the cooling timescale.  
Changes in other parameters can lead to faster cooling that 
may counteract the reduction in cooling at higher velocities. 

\begin{figure}[ht]
\centerline{
\includegraphics[width=1.0\columnwidth]{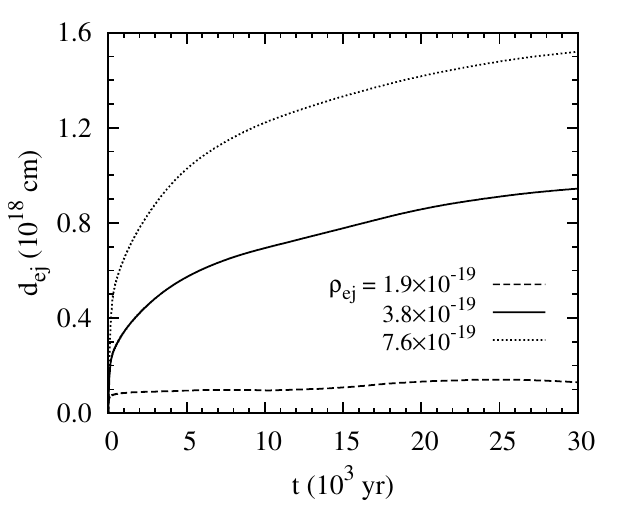}
}
\caption[]{Variation of ejecta delivery distance $d_{\rm ej}$ with time,
for 3 different values of the clump density.  The three curves correspond to 
$\rho_{\rm ej} = 1.9 \times 10^{-19} \, {\rm g} \, {\rm cm}^{-3}$ (dotted),
$3.8 \times 10^{-19} \, {\rm g} \, {\rm cm}^{-3}$ (solid),  
and $7.6 \times 10^{-19} \, {\rm g} \, {\rm cm}^{-3}$ (dashed).
Other clump parameters are set to their fiducial values: 
$V = 2000 \, {\rm km} \, {\rm s}^{-1}$, $R = 5 \times 10^{15} \, {\rm cm}$.}
\label{density}
\end{figure}  

\subsubsection{Effect of Clump Density}

The masses and radii of ejecta clumps are uncertain.  
The clumps in the Cas A supernova remnant appear to have radii as small as
$0.5 - 1 \times 10^{16} \, {\rm cm}$, comparable to (but slightly smaller than)
the radii $\approx d / 300$ inferred by Ouellette et al.\ (2010) to match 
(the largest) homologously expanding clumps in numerical simulations and 
observed supernova remnants.
The clump masses are difficult to ascertain, but the numerical simulations of 
Kifonidis et al.\ (2003, 2006) suggest masses $\sim 10^{-4} \, M_{\odot}$.
Each of these estimates is associated with uncertainties of a factor of a few.
For a mass $1 \times 10^{-4} \, M_{\odot}$ and radius $5 \times 10^{15} \, {\rm cm}$,
the clump density is $3.8 \times 10^{-19} \, {\rm g} \, {\rm cm}^{-3}$, but this
quantity is necessarily uncertain as well. 
To assess the effects of varying mass, density and radius, we choose to vary
the two parameters, density and radius.
In this subsection we assess the effect of varying density, and we consider 
clump densities $\rho_{\rm ej} = 1.9 \times 10^{-19}$, $3.8 \times 10^{-19}$, 
and $7.6 \times 10^{-19}$ g cm$^{-3}$. 
Because the radius is held fixed, the mass of each clump increases in proportion
to the density. 

Fig.~\ref{density} plots the results for $d_{\rm ej}$ from simulations with 
these three different clump densities (cases 7, 3 and 8). 
Other parameters are fixed at their fiducial values ($V = 2000 \, {\rm km} \, {\rm s}^{-1}$,
$R = 5 \times 10^{15} \, {\rm cm}$, density of the molecular cloud gas 
$= 3.3 \times 10^{-20} \, {\rm g} \, {\rm cm}^{-3}$). 
The ability of supernova ejecta to penetrate into the cloud is seen to increase 
significantly with increasing density (and mass).
For a clump density $\rho_{\rm ej} = 1.9 \times 10^{-19}$ g cm$^{-3}$, only
6 times denser than the molecular gas, the ejecta reach a negligible distance 
($\sim 10^{17}$ cm).
Doubling the density to the canonical value,
$\rho_{\rm ej} = 3.8 \times 10^{-19} \, {\rm g} \, {\rm cm}^{-3}$,
allows the density to penetrate to depths $> 9 \times 10^{17} \, {\rm cm}$,
9 times greater.  
A further doubling of the clump density, to 
$\rho_{\rm ej} = 7.6 \times 10^{-19} \, {\rm g} \, {\rm cm}^{-3}$,
increases the penetration depth, but by only a factor of 1.6, to 
$1.5 \times 10^{18} \, {\rm cm}$.
It appears that a critical threshold of clump density exists,
below which the penetration of ejecta into the molecular cloud is inefficient.
For $V = 2000 \, {\rm km} \, {\rm s}^{-1}$ that threshold is at or just 
below the canonical density 
$\rho_{\rm ej} = 3.8 \times 10^{-19} \, {\rm g} \, {\rm cm}^{-3}$.
We have also conducted a set of runs with varying density but with 
$V = 1000 \, {\rm km} \, {\rm s}^{-1}$ (cases 11, 2 and 12).
In these runs the penetration depth is insensitive to $V$ at higher densities,
but the case with $\rho_{\rm ej} = 1.9 \times 10^{-19} \, {\rm g} \, {\rm cm}^{-3}$
shows ejecta reaching significantly greater depths than in the 
$V = 2000 \, {\rm km} \, {\rm s}^{-1}$ case (run 7).
This implies that if a threshold exists, it is at lower density for the 
$V = 1000 \, {\rm km} \, {\rm s}^{-1}$.

The existence of such a critical density is understandable in the context of 
the early phase of evolution before radiative cooling becomes efficient. 
During this phase, the high pressure gradient in the postshock region decelerates 
the ejecta velocity around the $y,z=0$ axis, and deflects the $x$-momentum into 
the lateral  direction.  
For a clump of smaller density, this effect is stronger and proceeds more rapidly.
A clump with $V = 2000 \, {\rm km} \, {\rm s}^{-1}$ and 
$\rho_{\rm ej} = 1.9 \times 10^{-19}$ g cm$^{-3}$ 
is slowed too quickly before a cooling timescale to produce a narrow channel
in the cloud, and resembles again the cases without cooling.
In contrast, a clump with $V = 1000 \, {\rm km} \, {\rm s}^{-1}$ and 
$\rho_{\rm ej} = 1.9 \times 10^{-19}$ g cm$^{-3}$ manages to cool more 
rapidly, allowing supernova material to reach greater depths. 

For densities above the threshold, the penetration depth $d_{\rm ej}$ could
be expected to scale with the momentum of the clump and be linear in 
$\rho_{\rm ej}$.  
In fact, we see this trend for the cases with $V = 2000 \, {\rm km} \, {\rm s}^{-1}$ 
and $\rho_{\rm ej} = 3.8 \times 10^{-19}$ g cm$^{-3}$ (case 3) and 
$\rho_{\rm ej} = 7.6 \times 10^{-19}$ g cm$^{-3}$ (case 8).
The clump that is twice as massive travels almost a factor of 2 greater
($d_{\rm ej}$ is 1.6 times larger).
Likewise, for the $V = 1000 \, {\rm km} \, {\rm s}^{-1}$ runs (cases 2 and 12),
the clump that is twice as massive travels a factor of 1.5 times farther,
broadly consistent with the clump following a ballistic trajectory governed
by momentum conservation.

\subsubsection{Effect of Clump Size}

\begin{figure}
\centerline{
\includegraphics[width=1.0\columnwidth]{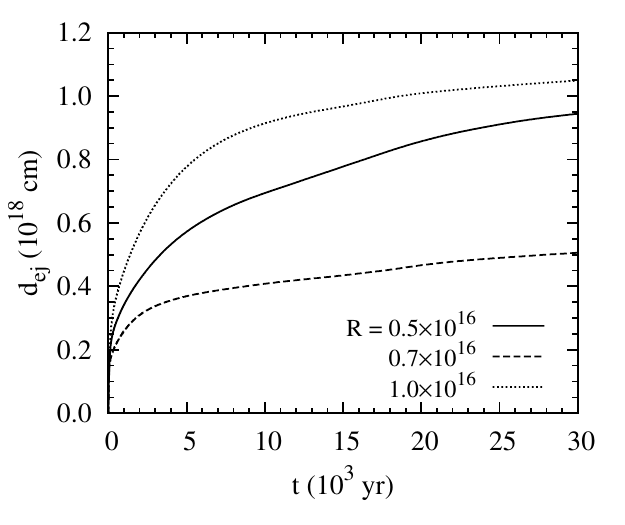}
}
\caption[]{Ejecta delivery distance $d_{\rm ej}$ as a function of time,
for three clump sizes, $R=0.5 \times 10^{16} \, {\rm cm}$ (solid),  
$0.7 \times 10^{16} \, {\rm cm}$ (dashed), and $1.0 \times 10^{16} \, {\rm cm}$
(dotted).  Other clump parameters are set to their fiducial values:
$V = 2000 \, {\rm km} \, {\rm s}^{-1}$, 
$\rho_{\rm ej} = 3.8 \times 10^{-19} \, {\rm g} \, {\rm cm}^{-3}$.}
\label{size}
\end{figure}  

As outlined above, the radii of the clumps is uncertain by factors of several.
In this subsection we assess the effect of varying the clump radius, and we
consider three values of the radius, $R = 5 \times 10^{15} \, {\rm cm}$,
$R = 7 \times 10^{15} \, {\rm cm}$ and $R = 1 \times 10^{16} \, {\rm cm}$
(cases 3, 9 and 10).
Smaller values are not ruled out by the optical images of the Cas A supernova
remnant (Fesen et al.\ 2001), but neither are they observed, and in any case 
the higher refinement levels necessary to model these scales are prohibitively 
expensive.
Because density is fixed at $\rho_{\rm ej} = 3.8 \times 10^{-19} \, {\rm g} \, {\rm cm}^{-3}$
in these runs, doubling the radius leads to a factor of 8 increase in mass, 
but only a factor of 2 increase in column density. 
In Fig.~\ref{size} we show how $d_{\rm ej}$ varies with time for the three
considered values of clump radius.

\begin{figure}
\centering
\centerline{
\includegraphics[width=.9\columnwidth]{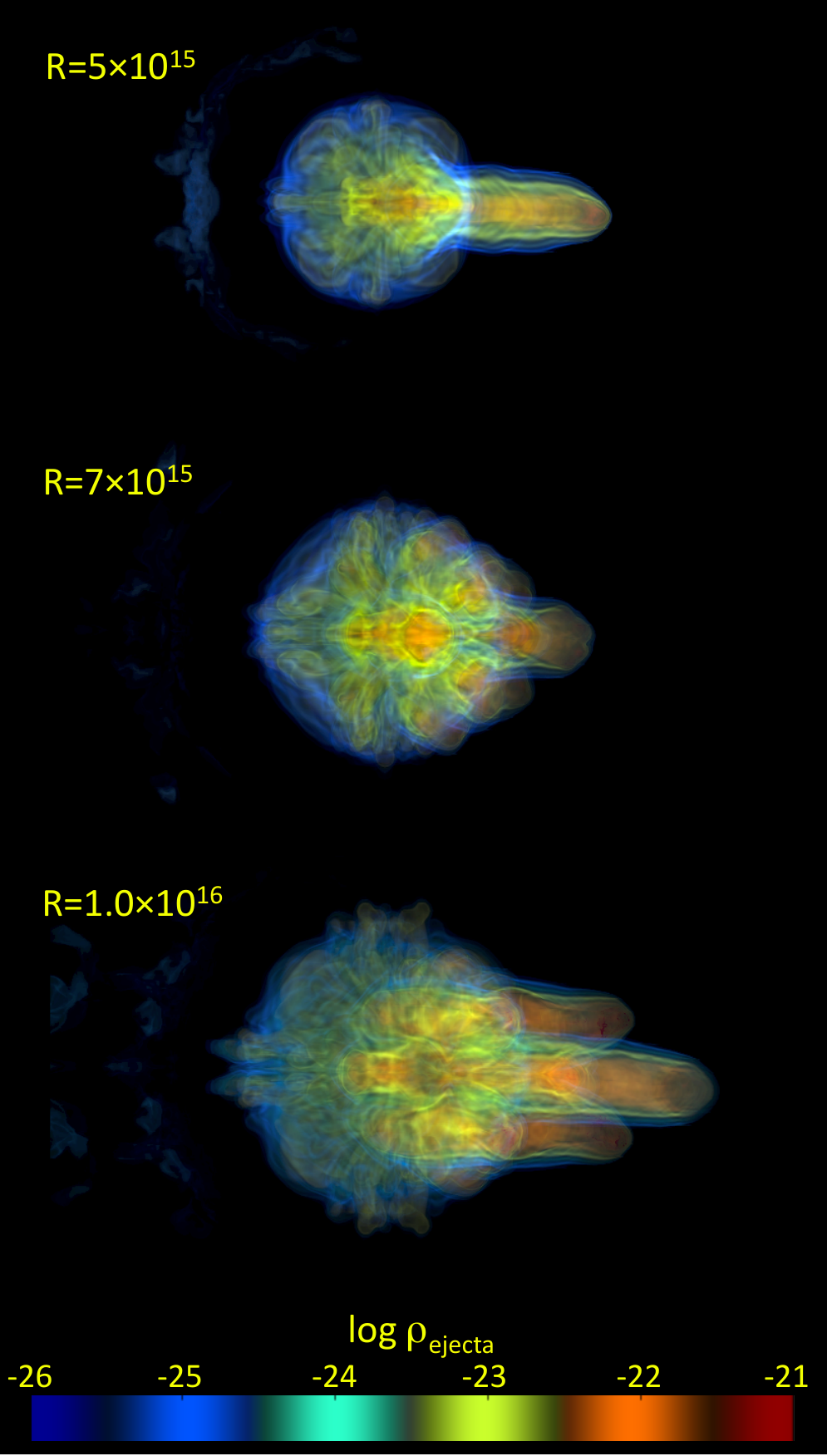} 
}
\caption[]{Three-dimensional rendering of spatial variation of ejecta density 
for fiducial parameters and three values of the clump radius.  The radii in
units of cm are labeled and correspond to cases 2, 13 and 14.
Larger clumps are seen to fragment more.}
\label{3dplot}
\end{figure}

The distance to which ejecta are delivered might be expected to increase with
the column density of the clump, and therefore its momentum per cross-sectional
area.
To some extent this trend of increasing $d_{\rm ej}$ with increasing $R$ is 
seen in Fig.~\ref{size}, in that the $R = 1 \times 10^{16} \, {\rm cm}$ 
case penetrates farther than the $R = 7 \times 10^{15} \, {\rm cm}$ case.
Confusing the issue, however, is the fact that $d_{\rm ej}$ for the 
$R = 5 \times 10^{15} \, {\rm cm}$ case is intermediate between the two
cases with larger $R$. This is because the delivery distance $d_{\rm ej}$ is controlled to
a large degree by fragmentation of the ejecta material, probably due to a 
cooling instability, that occurred at $\simeq 100 \, {\rm yr}$. 
We observed that a cone-like structure formed around the penetration axis after 
the fragmentation. 
Compared to the narrow channel formed in the case with $R= 5 \times 10^{15}$ 
(see Fig.\ 3), the cone-like structure has a larger cross section, and thus would 
need to sweep up more molecular gas to deliver the ejecta to the same distance. 
Fragmentation and production of a cone-like structure around the penetration axis
also occurred for the clump of the largest size, $R = 1 \times 10^{16} \, {\rm cm}$, 
but in this case a significant fraction of the ejecta remained concentrated along
the $y,z=0$ axis, opening a narrow channel into the molecular cloud. 
Because ejecta delivery through the channel is so efficient, the production of
a more well defined channel led to a greater delivery distance $d_{\rm ej}$.
In tandem, the fraction of ejecta remaining in the molecular cloud is seen to 
increase with increasing $R$. 
The effects of clump radius $R$ on the tendency of the clump to fragment are 
clearly seen in Fig.~\ref{3dplot}, which presents three-dimensional renderings
of the ejecta density (gas density times concentration field) for our fiducial 
parameters but three values of the clump radius. 

Repeating these calculations for a lower clump velocity, 
$V = 1000 \, {\rm km} \, {\rm s}^{-1}$ (cases 2, 13 and 14),
we find that the delivery distance tends to increase monotonically with 
increasing $R$. 
In these runs, the clumps remained coherent and did not fragment.
The trends of $d_{\rm ej}$ with $R$ suggest that the delivery distance is 
controlled by the momentum per cross-sectional area of the clump, scaling
roughly linearly with $R$. 
As before, the fraction of ejecta remaining in the cloud increases with 
increasing $R$, and is generally higher.
Combined with the above results, delivery of ejecta appears to become more 
complicated when the clump size is large, and is sensitive to the manner in 
which the clump fragments, perhaps by a thermal instability.

We attribute the fragmentation we observe to a cooling instability that occurs  
when the cooling function is such that compression leads to runaway cooling
(Field 1965; Field et al.\ 1969). 
This has the effect of increasing density perturbations in the 
post-shock region, leading to fragmentation of the ejecta.
This fundamentally changes its interaction with the surrounding material as 
it moves through the molecular cloud.

The sensitivity of the delivery distance to fragmentation begs the question 
of whether the ejecta clump would fragment for the fiducial parameters 
(including $R = 5 \times 10^{15} \, {\rm cm}$) if a higher numerical resolution 
were employed. 
It is not possible to completely rule out the existence of instabilities on 
smaller scales, but we did not observe the fragmentation behavior during our
high-resolution run with twice the numerical resolution (case 16; see \S 4.5).
We speculate that $R = 5 \times 10^{15} \, {\rm cm}$ and $V = 2000 \, {\rm km} \, {\rm s}^{-1}$
might represent a threshold case. 

To summarize, the trends with density, radius and mass,
the distance reached by supernova material in the molecular
cloud scales with the momentum per cross-sectional area of the clump, i.e.,
as $\rho_{\rm ej} R$ or $M / R^2$.
Thus, for a fixed density, larger radii and mass lead to larger $d_{\rm ej}$ although this
general trend is complicated by the possibility that clumps can
fragment, reducing $d_{\rm ej}$ because the ejecta are spread
out over a larger cross-sectional area, slowing them more quickly. 
On the other hand, increasing the density at a fixed radius increases the penetration
distance both because it increases the initial momentum per cross until area, 
and because it reduces the cooling time, reducing lateral  spreading in the early
stages of the collision.  
The fraction of ejecta remaining in the molecular cloud at 30,000 years 
tends to be large ($\gtsimeq 80\%$) when $d_{\rm ej} > 0.1 \, {\rm pc}$.

\subsection{Numerical Convergence}

\begin{figure}[ht]
\centerline{
\includegraphics[width=1.0\columnwidth]{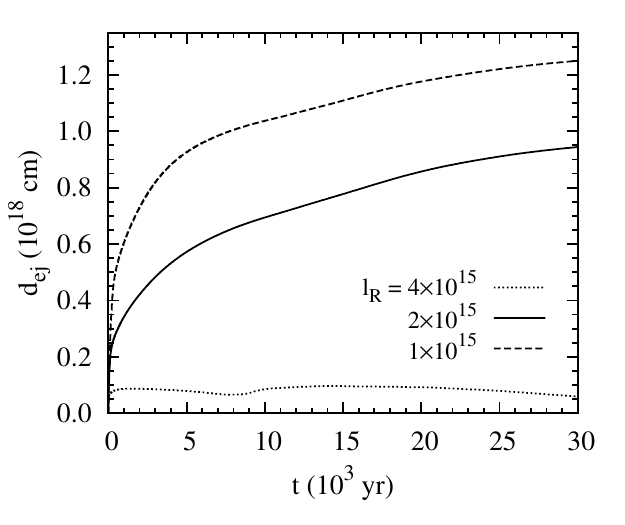}
}
\caption[]{Ejecta delivery distance $d_{\rm ej}$ as a function of time, for three
different values of the effective numerical resolution, $l_{\rm R} = 4 \times 10^{15} \, {\rm cm}$
(dotted), $2 \times 10^{15} \, {\rm cm}$ (solid), and $1 \times 10^{15} \, {\rm cm}$ (dashed). 
Other clump parameters were set to their fiducial values: $V = 2000 \, {\rm km} \, {\rm s}^{-1}$, 
$\rho_{\rm ej} = 3.8 \times 10^{-19} \, {\rm g} \, {\rm cm}^{-3}$, and $R = 5 \times 10^{15} \, {\rm cm}$.}
\label{3Dresolution}
\end{figure}

Finally,  the sensitivity of our calculated results to instabilities acting on small 
scales strongly motivates a study of whether we have achieved or are 
approaching numerical convergence.
In Fig.~\ref{3Dresolution} we track the delivery distance of supernova 
material as a function of time, for three effective grid resolutions, 
$l_{\rm R} = 4 \times 10^{15} \, {\rm cm}$ (case 15), 
$l_{\rm R} = 2 \times 10^{15} \, {\rm cm}$ (case 3), 
and $l_{\rm R} = 1 \times 10^{15} \, {\rm cm}$ (case 16).
Other parameters were held at their fiducial values. 
It is clearly seen that the clump is essentially not resolved for 
$l_{\rm R}= 4 \times 10^{15}$, in which case the entire clump is concentrated
in a few grid zones. 
In this case the ejecta material spreads rapidly in the lateral  direction
due to numerical diffusion, effectively stopping the clump. 
With increasing resolution this effect decreases, and we observe the root-mean-square
displacement of the ejecta material to decrease.
The greater concentration of ejecta around the penetration axis leads to 
an increase of $d_{\rm ej}$. 
As the resolution increases from $2 \times 10^{15} \, {\rm cm}$ to
$1 \times 10^{15} \, {\rm cm}$ (so that 10 zones are spread across the
diameter of the clump), $d_{\rm ej}$ increases by 30\%, almost entirely
due to physical effects occurring in the first $< 10^3$ years of the simulation.
After this initial stage, during which the clump moves about
$\approx 3 \times 10^{17} \, {\rm cm}$ (for $l_{\rm R} = 2 \times 10^{15} \, {\rm cm}$) or
$\approx 6 \times 10^{17} \, {\rm cm}$ (for $l_{\rm R} = 1 \times 10^{15} \, {\rm cm}$),
the clump moves an additional $\approx 6 \times 10^{17} \, {\rm cm}$ for both
resolutions. 

These results indicate that numerical convergence has not yet been achieved,
at least during the initial ($< 10^3 \, {\rm yr}$) stages.
Increasing numerical resolution introduces several competing effects.
On the one hand, higher resolution prevents the numerical diffusion that can 
artificially spread the ejecta.
On the other hand, increasing numerical resolution prevents numerical 
suppression  of physical instabilities like fragmentation that would actually spread the
ejecta in nature.  
Strict numerical convergence requires these instabilities to dominate and
reach a saturated state, which may not be easily achieved.
Another effect of higher resolutions is that density and temperature
fluctuations are better captured. 
Because of the quadratic dependence of cooling rate on density, higher resolutions
can lead to a faster overall cooling rate in the postshock region, which
would have the effect of decreasing $d_{\rm ej}$. 

Unfortunately, simulations with higher grid resolution are not possible in the context
of the current study. Because FLASH is an explicit code subject to the Courant-Friedrichs-Lewy
(CFL) stability criterion, doubling the grid resolution in a 3D simulation effectively
increases the number of grid zones by a factor of $\approx 2^3$ and the number of 
timesteps by a factor of 2, meaning over an order-of-magnitude increase in 
computational time. 
Although these effects can be ameliorated somewhat by adopting more aggressive
derefinement criteria and further restricting the computational domain size, we
expect that  increasing the resolution to $l_{\rm R} = 5 \times 10^{14} \, {\rm cm}$ 
would require $\approx 2$ million CPU hours per run, which is beyond our capabilities
here.


\begin{figure}[ht]
\centerline{
\includegraphics[width=1.0\columnwidth]{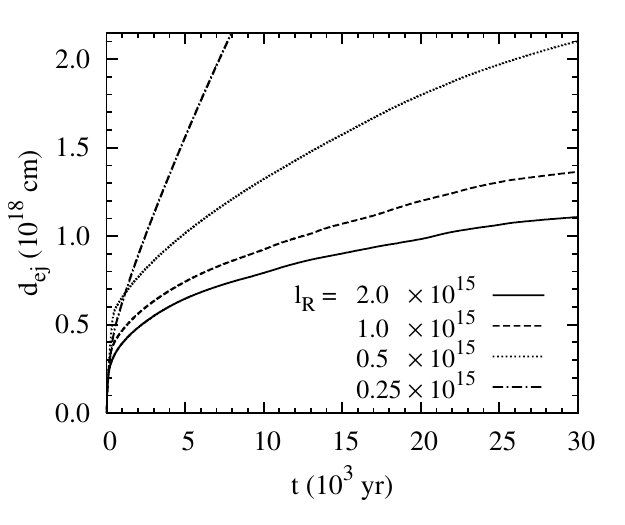}
}
\caption[]{Ejecta delivery distance $d_{\rm ej}$ as a function of time, from 2D simulation 
runs with cylindrical coordinates, for four different values of the effective numerical 
resolutions, $l_{\rm R} = 2 \times 10^{15} \, {\rm cm}$ (solid), 
$1 \times 10^{15} \, {\rm cm}$ (dashed), $0.5 \times 10^{15} \, {\rm cm}$ (dotted),
$0.25 \times 10^{15} \, {\rm cm}$ (dot-dashed).
Other parameters are fixed at their fiducial values.}
\label{2Dresolution}
\end{figure} 

In an attempt to study the problem at even higher grid resolutions, we carried out a
series of 2D (cylindrical geometry) runs with our fiducial values of $V$, $\rho_{\rm ej}$
and $R$. 
In Fig.~\ref{2Dresolution} we show the variation of $d_{\rm ej}$ with time for four
numerical resolutions, $l_{\rm R} = 2 \times 10^{15} \, {\rm cm}$, $1 \times 10^{15} \, {\rm cm}$, 
 $0.5 \times 10^{15} \, {\rm cm}$, and $0.25 \times 10^{15} \, {\rm cm}.$ 
 Note that while such 2D runs are able to 
achieve higher grid resolutions, it also
likely that they suppress instabilities present in the 3D run that laterally spread
out the clump material, causing it to slow. 
A comparison of Fig.~\ref{2Dresolution} with Fig.~\ref{3Dresolution} reveals that
$d_{\rm ej}$ in the 2D runs were typically 10-15\% greater than in the corresponding
3D runs at a given numerical resolution. 
This is likely due to the imposition of axisymmetry, which suppresses some
KH or other unstable modes that would spread the material, and by the restriction that 
material must remain centered on the penetration axis. 
For example, the fragmentation we observed in the 3D cases with $R = 7 \times 10^{15} \, {\rm cm}$
was not observed in the corresponding 2D run. 
These factors suggest that the 2D runs are an imperfect analog to the 3D runs and are
not well suited for studying numerical convergence.  

Adding to the peculiarities of the 2D simulations, we do not observe a converging 
trend in the variation of $d_{\rm ej}$ with increasing numerical resolution.
As the grid resolution is decreased from $2 \times 10^{15} \, {\rm cm}$
to $1 \times 10^{15} \, {\rm cm}$, the delivery distance increases by 25\%.
As the resolution is decreased another factor of 2, to $0.5 \times 10^{15} \, {\rm cm}$,
the delivery distance increases by 50\%.  
It is not clear whether these trends are attributable to physical effects or
numerical artifacts.

A previous study by Mac Low \& Zahnle (1994) of a similar problem, the interaction 
of cometary fragments with the Jovian atmosphere, showed that the numerical convergence 
required the smallest grid scale to be 4\% of the comet diameter, i.e., at least 
25 computational zones across the projectile were required. 
This problem differs from ours in that radiative cooling was not included (an
adiabatic gas with $\gamma = 1.2$ was assumed), and because the density of the
ambient gas varies with depth, unlike our molecular cloud. 
Indeed, even at the highest resolution we achieved in 2D ($l_{\rm R} = 0.25 \times 10^{15} \, {\rm cm}$,
or 40 zones across the clump, we did not observe any converging trends.
We infer that this lack of convergence is due to the fact that 
in 2D, physical instabilities that might spread the clump laterally are suppressed,
and numerical diffusion dominates the lateral spreading; and we infer that 
numerical diffusion is still significant even at our highest resolution. 

In summary, we conclude that 2D simulations fundamentally differ from the 3D simulations,
and do not provide a useful comparison.
Our 3D runs show some tendencies toward convergence but are not numerically converged.
The problem of injection of supernova clumps into a molecular cloud is a numerical challenge.
Even for the AMR code FLASH, the simultaneous need to resolve the clumps on small scales 
$< 10^{15} \, {\rm cm}$ and to model their behavior on large scales ($> 10^{18} \, {\rm cm}$)
is difficult to meet. 
Adding to the difficulty is that cooling can be effective, and via cooling instabilities 
can drive the shocked gas to collapse to smaller scales.
On the other hand, many aspects of our 3D runs appear to be robust. 
Ejecta clumps with fiducial parameters do tend to be injected into the molecular cloud,
reaching depths  $\sim 10^{18} \, {\rm cm},$  with a high fraction
$\gtsimeq 80\%$ of the ejecta remaining in the molecular cloud even after 30,000 years.

\section{Discussion}

\subsection{Summary}

The numerical simulations described above represent the first numerical study 
of clumpy supernova ejecta interacting with molecular gas at the periphery of 
an H {\sc ii} region. 
We assumed typical distances from the supernova of about 2 pc, similar to the distances
of ejecta from the explosion center in Cas A, and comparable to
the distance an ionization front is inferred to propagate before supernova occurs.
Guided by the approximate models of Ouellette et al.\
(2010) and by observations of the Cas A supernova remnant, we assume
radii $\approx \times 10^{15} \, {\rm cm}$, masses $\approx 1 \times 10^{-4} \, M_{\odot}$,
and densities $\approx 3.8 \times 10^{-19} \, {\rm g} \, {\rm cm}^{-3}$. 
Furthermore, we adopted a fiducial ejecta velocity $= 2000 \, {\rm km} \, {\rm s}^{-1}$, and the 
molecular gas density was fixed at $n_{\rm H2} \approx 10^{4} \, {\rm cm}^{-3},$
including a cooling function appropriate to shocked optically-thin gas.  

With these parameters, numerical resolution is a real concern.
Metrics like the mean distance traveled by ejecta 
after a stopping time (i.e., $d_{\rm ej}$ at 30,000 years)
are sensitive to physical conditions in the first 1000 years of the 
interaction and vary non-monotonically as the numerical resolution increases.
Convergence is difficult to achieve because of the very large span of length scales in 
the problem: clumps travel $\sim 10^3$ times their own diameter, and fragment by KH and 
possibly cooling instabilities, into even smaller scales.
Unfortunately, higher numerical resolution is infeasible for this study, as each run 
consumed several hundred thousand CPU-hours.
Turbulence, which is not included in these runs, may ameliorate these problems 
somewhat by introducing a lower limit to the size of coherent fragments.

Despite the lack of numerical convergence, certain trends in the data appear 
to be robust.
Under the right conditions, $\gtsimeq 80-90\%$ of the clump material 
is injected to mean depths $\approx 0.3 \, {\rm pc}$ 
and remains in the molecular cloud.
The conditions under which ejecta remain in the cloud appear to hinge entirely 
on the cooling timescale.
If cooling is not sufficiently rapid, the post-shock pressure builds to the point that the 
bulk of the ejecta is expelled from the molecular cloud.
Efficient injection requires a cooling timescale not much greater than the dynamical
timescale, $\sim 100 \, {\rm yr}$.
The cooling timescale decreases in inverse proportion to the post-shock density, 
and a threshold density exists for injection of material, which is very roughly 6 times the density 
of the molecular gas. 
The cooling timescale also increases sensitively with post-shock temperature,
and therefore shock speed.
An optimal ejecta velocity $V \approx 1000 - 2000 \, {\rm km} \, {\rm s}^{-1}$
exists for injection, and the depth of delivery, $d_{\rm ej},$ and the fraction injected, $f_{\rm inj},$ decrease with 
increasing $V$. 

Another robust trend is that if cooling is effective, $d_{\rm ej}$
appears to correlate with the clump's momentum per unit area\footnotemark\footnotetext{The momentum per unit area 
has been found to be a crucial quantity in a different context by Foster and Boss (1996). 
In a study of the interaction of stellar ejecta with molecular cloud cores, they showed that the 
incident momentum of the ejecta plays a important role in determining whether  the interaction 
leads to collapse or destruction of the cloud core.}.
Clumps with higher $\rho_{\rm ej} R$ generally travel farthest, but 
$d_{\rm ej}$ does not monotonically increase with this quantity because
instabilities can cause the clump to laterally spread, decreasing
its momentum per area. 
These instabilities manifest themselves more prominently when the radius
of the clump is increased and is therefore better resolved. 
In all cases where the clump penetrates beyond $\gtsimeq 10^{17} \, {\rm cm}$,
 $\gtsimeq 80\%$ of the clump material remains in the molecular
cloud at late times, and $f_{\rm inj}$ increases with $d_{\rm ej}$. 

Our ability to numerically model the full range of parameters relevant to the 
supernova injection problem is incomplete.
For example, clumps can be smaller (below the imaging resolution of the {\it Hubble
Space Telescope}) and denser than we accounted for, and many clumps in the Cas A supernova 
remnant are likely to be traveling at speeds $V \approx 6000 \, {\rm km} \, {\rm s}^{-1}$ 
(Fesen et al.\ 2001).
Smaller clumps and faster shock speeds are very difficult to numerically compute; 
because of the limitations of the CFL condition, even a factor of 2 increase in 
resolution requires an order of magnitude more computing time.
As well, we did not vary the density of gas in the molecular cloud. 
Gas at the tops of the pillars in M16 
has been shocked by the advancing D-type ionization front, and
is characterized by densities
$\gtsimeq 10^5 \, {\rm cm}^{-3}$, an order of magnitude higher than the densities we assumed.
Density inhomogeneities such as star-forming cores within the molecular cloud would 
affect the propagation of ejecta material as well, and  turbulent motions  which must
be present in the molecular cloud would affect the diffusion of material and the minimum lengthscales of
coherent structures.

Despite these limitations, we have gained enough insight from our parameter studies 
to predict how injection would proceed under different scenarios.
One relevant set of parameters might be high-velocity ($6000 \, {\rm km} \, {\rm s}^{-1}$)
clumps encountering denser ($n_{\rm H2} = 10^5 \, {\rm cm}^{-3}$) molecular gas.
A shock speed that is a factor of 3 higher than our canonical value suggests a 
postshock temperature an order of magnitude higher ($T \propto V^2$), and a cooling
timescale about a factor of 3 longer (see Fig.~1). 
On the other hand, the higher density of gas in the molecular cloud leads to postshock
densities an order of magnitude greater, and a cooling timescale a factor of 10 smaller.
This means cooling timescales are likely to be sufficiently
short to allow efficient injection of clumpy supernova material into molecular gas that has already been
shocked by a D-type ionization front.

While much work remains to be done, our initial investigations clearly indicate that supernova clumps can be
 injected efficiently in to molecular material in many cases.
When the ejecta finally come to rest, a large fraction of the clump 
mass will remain in the molecular cloud, mixing into material that is in the midst of 
collapsing to form new Sun-like stars.

\subsection{Impact on Solar System Isotopic Anomalies}

To determine the degree to
which a forming solar system is contaminated by supernova material,
we assume there are $N \sim 10^4$ clumps of mass $M \sim 10^{-4} \, M_{\odot}$, 
so that $N M = M_{\rm ej}$, the mass of the ejecta that is not ejected
isotropically.  
Implicitly neglecting density variations within the molecular cloud,
we assume that the periphery of the H {\sc ii} region traces a sphere
of radius $r$ centered on the supernova, consistent with the assumption that
the supernova progenitor was the dominant source of the ionizing photons that 
carved out the H {\sc ii} region. 
As discussed in \S 2.1, based on a main-sequence lifetime for
the progenitor of 5 Myr and an ionization front that advances at 
$0.4 \, {\rm km} \, {\rm s}^{-1}$, we adopt a value $r \approx 2 \, {\rm pc}$.
On average, the cross-sectional area of the molecular cloud that is 
associated with each clump is $4\pi r^2 / N$ $\sim (2 \times 10^{17} \, {\rm cm})^2$
or $\sim (0.06 \, {\rm pc})^2$ for our adopted parameters, such that
clumps will be separated by $\sim 4 \times 10^{17} \, {\rm cm}$.

The separation between clumps is surprisingly close to the width of the channel 
that is carved out in our fiducial 
runs, $\approx 2 \times 10^{17} \, {\rm cm}$, which is not significantly wider than 
the distribution of ejecta (Fig.~3).
This suggests that lateral mixing may be rapid enough to contaminate
the gas between channels.
We do not explicitly model the turbulence that would effect this mixing, but
we can estimate the mixing timescale by assuming that turbulence operates at least as
effectively as in the pre-shocked molecular cloud.
The turbulent mixing timescale at a lengthscale $l$ scales as $l / \delta v(l)$,
where $\delta v(l)$ is the amplitude of the velocity fluctuations on the scale $l$
(Pan \& Scannapieco 2010).
On the scale of the channel separation, $\sim 0.1 \, {\rm pc}$, we estimate 
$\delta v(l) \approx 1 \, (l / {\rm pc})^{0.4} \, {\rm km} \, {\rm s}^{-1}$
using the Larson scaling law (Larson 1981).
Thus the mixing timescale across 0.1 pc would be $2 \times 10^{5} \, {\rm yr}$.
This should be viewed as an upper limit for mixing, because of several
factors that would increase turbulence and $\delta v(l)$.
For example, the interaction of the underlying turbulent gas with the shocks
created by the clump may increase the turbulent velocity fluctuations 
(e.g., Lee et al.\ 1997).
The mixing timescale of $\sim 10^5 \, {\rm yr}$ is slightly longer than the 
duration of our simulations, but interestingly are comparable to the free-fall
timescale ($\sim 3 \times 10^5 \, {\rm yr}$) on which this molecular gas will form 
protostars.
Mixing across multiple channel separations (i.e., 0.3 pc), however, takes significantly
longer.
Thus, for the purposes of estimating the magnitude of isotopic anomalies, we assume
that the gas between channels is contaminated effectively by the ejecta deposited in
the nearest few channels.

Assuming effective mixing, the volume of molecular gas that is associated with a single 
clump is $4\pi r^2 \, d_{\rm sh} / N$. 
If the ejecta mix evenly throughout this entire volume,  the fraction of the
mass that comes from the supernova would then be
\[
f_{\rm cont} \approx \frac{ M }{ \rho_{\rm MC} \, 4\pi r^2 \, d_{\rm sh} / N}
 = 1 \times 10^{-4} \, 
   \left( \frac{M_{\rm ej}}{2 \, M_{\odot}} \right) \, 
   \left( \frac{ r }{ 2 \, {\rm pc} } \right)^{-2} \,
\]
\begin{equation}
   \times
   \left( \frac{ d_{\rm sh} }{0.5 \, {\rm pc}} \right)^{-1}
   \left( \frac{ \rho_{\rm MC} }{ 4.7 \times 10^{-20} \, {\rm g} \, {\rm cm}^{-3} } \right)^{-1}.
\end{equation} 
Here we must interpret $M_{\rm ej}$ as the total mass of clumpy supernova material, as 
isotropic ejecta do not inject efficiently. This is the average concentration of supernova 
material in the molecular gas, up to a depth of about 0.5 pc.  
Clearly, the physical parameters of individual ejecta clumps affect 
$f_{\rm cont}$ only through $d_{\rm sh}$, whose dependence on these parameters 
has been studied in \S 4.4. The fraction $f_{\rm cont}$ is likely insensitive to $\rho_{\rm MC}$. 
This is because the penetration of clumps into the molecular cloud, when successful, 
is limited by the momentum of the clumps, and thus the product of $d_{\rm sh}$ and $\rho_{\rm MC}$ 
is expected to be roughly constant. 
Note that this is an {\it average} concentration of all the ejecta lying within 0.5 pc
of the ionization front, and
higher concentrations are possible in smaller fractions of the volume.
Note also that the average concentration of supernova material is lower at greater 
distances, but still substantial.  
If the ionization front were at 4 pc instead of 2 pc, for example, the concentration would be 
reduced by less than a factor of 4, because the
clump would have expanded, lowering its column density and reducing $d_{\rm sh}$. 
The point is that even at a different distance the molecular gas still would be 
robustly contaminated by the supernova at a significant level. 

We note again that ongoing star formation is observed in the molecular gas at the edges of of H {\sc ii}
regions, probably triggered by the shocks driven a few tenths of a pc in advance of the 
ionization front (e.g., Snider et al.\ 2009 and references therein; see also \S 2.1).
There is the additional intriguing possibility that the shocks propagating through the 
molecular cloud, driven by the clumps themselves, could trigger new star formation.
This star formation is expected to take $10^{4} - 10^{5} \, {\rm yr}$, based on 
the evolutionary state of protostars that are uncovered by ionization fronts 
(Hester et al.\ 2004; Hester \& Desch 2005; Snider et al.\ 2009).
This is comparable to the mixing timescale derived above, suggesting that 
each protostar is likely to acquire material from just one or a few clumps, and
incorporation of this material is likely to take place shortly
before or during star formation. 
Thus, provided the Sun formed at the periphery of an H {\sc ii} region,
it is likely to have incorporated supernova material 
from a single small region or mixture of a few small regions from the nearby supernova. 
Adopting a mixing ratio $\approx 10^{-4}$, the abundance of an element or isotope in 
the solar nebula can be determined if the composition of that clump can be 
constrained. 
Obviously it is possible to pick any number of small regions within a ``prompt" supernova of 
any arbitrary mass $\gtsimeq 40 \, M_{\odot}$,
and a full exploration of the problem is not possible here, although we can
make general estimates.

For example,  the total amount of ${}^{26}{\rm Al}$ that is produced in a 
$40 \, M_{\odot}$ progenitor is $\approx 1.5 \times 10^{-5} \, M_{\odot}$ 
(Ellinger et al.\ 2010).
This implies that after mixing, $1 \, M_{\odot}$ mass of gas will contain about
$1.5 \times 10^{-9} \, M_{\odot}$ of ${}^{26}{\rm Al}$.
This is to be compared to the mass of ${}^{27}{\rm Al}$ in the solar system,
which is about $6.7 \times 10^{-5} \, M_{\odot}$ using the abundances of Lodders (2003).
This estimate immediately suggests that within the contaminated 
portions of the molecular cloud,
${}^{26}{\rm Al} / {}^{27}{\rm Al} \sim 2 \times 10^{-5}$,
{\it on average}.
This is remarkably comparable to the initial ${}^{26}{\rm Al} / {}^{27}{\rm Al}$ ratio
$\approx 5 \times 10^{-5}$ inferred for the solar nebula (MacPherson et al.\ 1995).

There are several factors that could cause the ${}^{26}{\rm Al} / {}^{27}{\rm Al}$ 
ratio to deviate significantly from this approximate average value. 
First, it is not certain that mixing within the molecular cloud following the injection 
can proceed to completion, so that regions near the channels
carved out by the ejecta might be over-enriched with respect to the surrounding gas.
Also, because the solar system could have been contaminated by any small region within 
the supernova material, which varies by an order of magnitude in its ${}^{26}{\rm Al}$ 
content.
For example, in the 1D calculations by Ellinger et al.\ (2010), 
the sub-explosive C-burning regions produced 
$\sim 10^{-5} \, M_{\odot}$ of ${}^{26}{\rm Al}$, despite making up a small fraction of the supernova
mass.
And in the 3D simulations by Ellinger et al.\ (2010), some 
$\sim 2 \times 10^{-5} \, M_{\odot}$ smoothed-particle-hydrodynamics (SPH) particles 
 contained $4.8 \times 10^{-10} \, M_{\odot}$ 
of ${}^{26}{\rm Al}$, yielding an even higher mass fraction of ${}^{26}{\rm Al}$. 
In fact, again assuming a mixing ratio $\sim 10^{-4}$, if only one of these ${}^{26}{\rm Al}$-rich clumps 
contaminated the solar nebula,
the initial ${}^{26}{\rm Al} / {}^{27}{\rm Al}$ ratio would be 
$7 \times 10^{-4},$ over ten times the observed value.
Currently, it is not possible to predict the initial abundance of ${}^{26}{\rm Al}$ any
better than this, but it is clear that if the solar nebula formed from molecular
gas contaminated by an ${}^{26}{\rm Al}$-rich clump, then its initial ratio
would be comparable to the value observed in meteorites.

Using the same example dataset of Ellinger et al.\ (2010) we can also estimate
the shifts in elemental and isotopic abundances of oxygen.
In their 1D model of a $40 \, M_{\odot}$ progenitor, the total ejected mass of O
(almost all ${}^{16}{\rm O}$) is $3.29 \, M_{\odot}$. 
Assuming a mixing ratio of $7 \times 10^{-5}$ implies that $2.3 \times 10^{-4} \, M_{\odot}$
of oxygen is injected, on average, into every $1 \, M_{\odot}$ mass of molecular gas
that will form a solar system.
This is to be compared to the mass of oxygen in the solar system, $6.7 \times 10^{-3} \, M_{\odot}$ 
(Lodders 2003), which implies that on average the late-forming stars at the edge of the H {\sc ii}
will see increases in their oxygen content by 3\%, although again some clumps will be
significantly more oxygen-rich than average.
The sub-explosive C burning zones in the 1D models themselves produced 
$2.47 \, M_{\odot}$ of oxygen (again, almost all ${}^{16}{\rm O}$),
despite their lower mass overall, suggesting that larger shifts in oxygen abundance,
potentially several tens of percent, are not unreasonable for some stars. 

The isotopic shifts in O associated with injection of supernova material also 
were considered by Ellinger et al.\ (2010).
Assuming that sufficient ${}^{26}{\rm Al}$ is injected into the forming solar
system to explain the meteoritic abundance
${}^{26}{\rm Al} / {}^{27}{\rm Al} = 5 \times 10^{-5}$, they found isotopic
shifts in oxygen could span a wide range of values. For the 1D models, a 
clump from the sub-explosive C-burning zones of a $40 \, M_{\odot}$ progenitor 
tends to inject nearly pure ${}^{16}{\rm O}$, dropping both the 
${}^{17}{\rm O} / {}^{16}{\rm O}$ and ${}^{18}{\rm O} / {}^{16}{\rm O}$ 
ratios in the solar system, equivalent to a decrease in $\delta^{17}{\rm O}$ by roughly 35 
permil in the cosmochemicalnotation. For this case there is little change in the ${}^{18}{\rm O} / {}^{17}{\rm O}$
ratio, but in 3D simulations, they found that many ${}^{26}{\rm Al}$ producing 
regions were significantly enhanced in ${}^{18}{\rm O}$ relative to ${}^{17}{\rm O}$, an effect that was especially strong 
in anisotropically exploding supernovae (Ellinger et al.\ 2011).
Thus, injection of enough ${}^{26}{\rm Al}$ to explain the meteoritic evidence could shift 
the ${}^{18}{\rm O} / {}^{17}{\rm O}$ ratio by a factor of 2, by increasing
$\delta^{18}{\rm O}$ by $> 1000$ per mil with little change in 
$\delta^{17}{\rm O}$. This injection of ${}^{18}$O into the solar system as it formed 
would decrease $\Delta {}^{17}O$ by a shift comfortably larger 
than the 300 permil shift inferred by Young  et al (2011).

In summary, exact shifts in elemental and isotopic abundances will depend on where within
the supernova the one or few clumps that contaminated the solar system came from,
so it is premature to try to predict the exact shifts. Nevertheless,
supernova contamination of molecular gas appears able to 
qualitatively explain the abundance of ${}^{26}{\rm Al}$ and the shifts in 
oxygen abundance and isotopic composition inferred for our solar system.

\subsection{Statistics of Supernova Contamination}

Injection of supernova material into an already-formed protoplanetary disk has
been critically examined by Gounelle \& Meibom (2007) and Williams \& Gaidos (2007).
The general point raised by these authors is that {\it recently formed} disks are
overwhelmingly likely to be several parsecs from the supernova progenitor, necessarily 
forming in the molecular gas at the periphery of the H {\sc ii} region.  
Looney et al.\ (2006) made similar points. 
Disks forming at $> 2 \, {\rm pc}$ from the supernova will intercept relatively
little ejecta, so that insufficient ${}^{26}{\rm Al}$ could be intercepted to
explain the meteoritic ratio.  
The meteoritic ratio ${}^{26}{\rm Al} / {}^{27}{\rm Al} = 5 \times 10^{-5}$,
if the disk intercepts isotropically exploding ejecta, requires the disk to
lie only $\sim 0.1 \, {\rm pc}$ from the supernova (Ouellette et al.\ 2005, 2007, 2010).
On the other hand, Ouellette et al.\ (2010) showed that a protoplanetary disk 
intercepting clumpy ejecta can receive much more material than a disk intercepting 
isotropic ejecta at the same distance.
For example, considering the model discussed above, $10^4$ clumps of mass
$2 \times 10^{-4} \, M_{\odot}$ each, with radii $d / 300$, will have a volume 
filling fraction $3.7 \times 10^{-4}$ and will be 2700 times denser than isotropically
expanding ejecta.  
Such a clump could intecept a protoplanetary disk at a distance of 2 pc
(at which point its radius would be $\sim 10^3 \, {\rm AU}$), and the disk would
receive as much supernova material as if it were exposed to isotropically expanding
ejecta at a distance of 0.04 pc.
If the clump samples a ${}^{26}{\rm Al}$-rich region within the supernova, the
disk could intercept even more ${}^{26}{\rm Al}$. 
On the other hand, the areal filling fraction of the ejecta clumps, at the boundary 
of the H {\sc ii} region, will be only 2.8\%, so only one in 36 disks at the time
of the supernova would encounter such clumps.  
Multiplying this fraction by the number of Sun-like stars forming disks late in the
evolution of an H {\sc ii} region, we estimate that $\approx 0.1 - 1\%$ of Sun-like
stars will encounter significant amounts of supernova material during the protoplanetary
disk stage.
This scenario, suggested by Chevalier (2000) and Ouellette et al.\ (2005, 2007, 2010),
remains a viable, but unlikely,  explanation for SLRs in the solar nebula.

In contrast, injection of supernova material into molecular gas, just as 
stars and planetary systems are forming, appears to be robust.
Essentially {\it all} stars forming late in the evolution of the star-forming
region will be contaminated by some type of clumpy supernova material.
The probability that a solar system would be contaminated is essentially the
fraction of stars forming in a cluster rich enough to have a star $\gtsimeq 40 \, M_{\odot}$
in mass (so that it explodes in $< 5 \, {\rm Myr}$), times the fraction of stars
that form in such a cluster after 5 Myr of evolution.
As outlined in \S 2.1, the first fraction is probably $\approx 75\%$. 
The second fraction depends on the rate of star formation and is related to the 
question of whether star formation is triggered. 

Multiple observations show that star formation is ongoing in H {\sc ii} regions,
even of ages 2-3 Myr (Palla \& Stahler 2000; Hester et al.\ 1996, 2004; Healy et al.\ 
2004; Sugitani et al.\ 2002; Snider 2008; Snider et al.\ 2009; Snider-Finkelstein 2009; 
Getman et al.\ 2007; Reach et al.\ 2009; Choudhury et al.\ 2010; Billot et al.\ 2010; 
Bik et al.\ 2010; Zavagno et al.\ 2010; Beerer et al.\ 2010; Comer\'{o}n \& Schneider 2011).
Snider (2009) examined the ages of recently formed stars in several H {\sc ii} regions
using combined {\it Spitzer Space Telescope} and {\it Hubble Space Telescope} data. 
The analysis of NGC 2467 in particular was presented by Snider et al.\ (2009),
who found that 30-45\% of the Sun-like stars in this H {\sc ii} region were 
triggered to form after the initial formation of the cluster and the most massive
stars, 2 Myr ago.  
This implies that if the rate of triggered star formation is constant in time,
and extends until the supernova explodes, at about 5 Myr of age, then $\sim 10\%$ of Sun-like stars 
in this cluster
would form in the last 1 Myr of star formation.
If the rate of triggered star formation scales as the area swept out by the 
ionization front and therefore as the square of the age of the cluster, $t^2$, 
then the fraction of stars forming in the last 1 Myr before the supernova could 
be as high as $\sim 40\%$. 
If approximately 75\% of all Sun-like stars form in a rich cluster with
a star that will go supernova within 5 Myr, and 10-40\% of those stars
form in the 1 Myr before the supernova, then the likelihood of a Sun-like
star forming from gas contaminated by ejecta from a recently exploded
supernova is on the order of 7-30\%.
This is a considerably higher probability than the $0.1 - 1\%$ probability
of injection of an ejecta clump into a protoplanetary disk 
(Ouellette et al.\ 2010).
More importantly, it suggests strongly that supernova contamination may be a 
 common and universal process. 

\subsection{Elemental Variability of Sun-Like Stars}

If supernova contamination is a common process, one would expect to see variations
in elemental abundances in spectra of Sun-like stars.
In fact, there is ample evidence from stellar spectra of planet-hosting stars
for variability in elemental abundances.  
Fischer \& Valenti (2005) surveyed 850 FGK-type stars that have Doppler 
observations sufficient to uniformly detect  all planets with radial velocity 
semiamplitudes $K > 30 \, {\rm m} \, {\rm s}^{-1}$ and orbital periods shorter 
than 4 yr. 
Among this sample they found variations of up to a factor two in [Na/Fe], 
[Si/Fe], [Ti/Fe], and [Ni/Fe] over the range -0.5$<$[Fe/H]$<$ 0.5, and no 
correlations between metallicity and orbital period or eccentricity. 
They concluded that host stars do not have an accretion signature that 
distinguishes them from non-host stars, and that host stars are simply 
born in higher-metallicity molecular clouds.  
Bond et al.\ (2008) analyzed elemental abundances for eight elements, 
including five heavy elements produced by the r- and s-processes, in 28 
planetary host dwarf stars and 90 non-host dwarf stars. 
They found elemental abundances of planetary host stars are only slightly 
different from solar values, while host stars are enriched over non-host 
stars in all elements studied, varying by up to a factor of two but with 
enrichments of 14\% (for O) and 29\% around the mean.  
Pagano et al. (2010) examined elemental abundances for 13 elements in 
52 dwarf stars in the solar neighborhood, and found the variations in 
C, O, Na, Al, Mg, Ca, and Ti to be about a factor of two around the mean 
at the 3 $\sigma$ level.

Supernova injection into the molecular cloud from which protostars are
forming remains a plausible mechanism for these variations, and may
contribute to the abundances observed in planet-hosting stars.
For example, as discussed in \S 5.2, if clumps from different regions 
of a supernova could be injected into a $1 \, M_{\odot}$ mass cloud core
as it was collapsing, at a mixing ratio $\approx 10^{-4}$ (see Eq.~1),
one could get variations of the order observed. 
For example, an oxygen-rich clump could have an increase of up to 30\%.

It is worth noting that if the variability in stellar elemental abundances
can be attributed to injection of supernova material, then observations
like those mentioned above could be used to assess whether an exoplanetary
system was contaminated by unobservable species.
These could include short-lived radionuclides like ${}^{26}{\rm Al}$,
which are long extinct in any system older than a few Myr, as well as  P, which is difficult to 
observe because it lacks optical transition lines.
Massive stars produce a number of isotopes within a given mass shell,
and these isotopic abundance ratios may be conserved within an
impinging clump if mixing within the supernova explosion is not large.
For example, Young et al.\ (2009) considered the co-spatial production 
of elements in supernova explosions, to find observationally detectable
proxies for enhancement of $^{26}$Al. 
Using several massive progenitor stars  and explosion models, they 
found that the most reliable indicator of  $^{26}$Al in 
an unmixed clump is a low S/Si ratio of $\sim$0.05. 
A clump formed from material within the O-Ne burning shell should be enriched 
in both $^{26}$Al and $^{60}$Fe (Timmes et al 1995, Limongi \& Chieffi 2006)
and the biologically important element P is produced at its highest
abundance in the same regions (Young et al 2009).
Even if these specific elemental ratios are not found, the supernova injection 
model broadly predicts  that species co-produced within supernovae will tend 
to show correlated excesses in stellar spectra.
Observations of elemental abundances from stellar spectra can be used to test
this hypothesis.

\subsection{Mixing on Galactic Scales} 

Finally, our results have implications for how the metals ejected by supernovae 
are released into the multiphase ISM, a question of key 
importance in understanding their turbulent mixing on  lengthscales 
$\approx 10 - 500 \, {\rm pc}$ (Roy \& Kunth 1995; Scalo \& Elmegreen 2004).  
Tenorio-Tagle (1996), for example, argued using simple estimates that metals are 
likely to be released directly into hot, thermalized superbubbles, which blow out 
of the galactic disk, only to cool and rain down later as metal-rich ``droplets" 
that are then broken apart by the RT instability.  
In this case, there would be a significant delay between metal production and 
enrichment, but after this delay metals would be deposited over large regions.  

A more detailed numerical study was carried out by de Avillez \& Mac Low (2002), 
who examined turbulent mixing in a multiphase ISM that was seeded with 
a scalar concentration field that varied on a fixed spatial  scale that was
uncorrelated with the locations of supernovae.   
They found that at early times the variance of the concentration 
decreased on a timescale that  was proportional to the lengthscale of the initial 
fluctuations, and they argued that the late-time evolution was 
largely independent of this lengthscale.   
At early times, these results can be understood as being controlled by the mixing of
metals in hot low-density environments, which occurs in any single temperature medium 
on a timescale set by the initial length fluctuations divided by the turbulent velocity 
(Pan \& Scannapieco 2010).  
On the other hand,  at late times the results might 
depend on the much slower process of mixing between hot and cold regions, which is set by the size of the cold 
clouds and their density contrast with the hot medium (e.g.\ Klein, McKee, \& Colella 
1994; Fragile et al.\ 2004).  
More recently, Ntormousi \& Burkert (2012) have emphasized the difficulty of mixing 
metals from the hot  gas into the colder ISM  out of which new molecular clouds 
form, arguing that the enrichment of the cold ISM will be 
delayed by at least a cooling time of the hot diffuse gas.

The mixing of clumpy supernova ejecta directly into molecular clouds, seen in our 
simulations, would completely circumvent this limiting step in galactic chemical 
evolution.  
While a fraction of the elements deposited by this mechanism would be locked into 
Sun-like stars formed in the wake of the D-type ionization front and supernova shock,  
at least as much enriched molecular material would be subsequently ionized and
launched into the low-density ($\approx 0.1 \, {\rm cm}^{-3}$), warm, 
ionized ($\approx 10^4 \, {\rm K}$) medium 
(e.g. Matzner \& McKee 2000).
The higher densities of this gas lead to shorter cooling times and higher
density contrasts, by orders of magnitude, greatly accelerating mixing, as compared
to superbubbles.
This process of warm-phase galactic enrichment merits further theoretical study 
and may be important in explaining the relative homogeneity of the Milky Way ISM on 
$\sim 100 \, {\rm pc}$ scales (Meyer et al.\ 1998; Cartledge et al.\  2006), 
as well as the low dispersions seen in massive stars in nearby galaxies 
(e.g.\ Kobulnicky \& Skillman 1996; 1997).

\subsection{Final Word}

Supernovae have long been implicated to explain stable isotope anomalies and 
the abundances of short-lived radionuclides in the early solar system.
As the Sun's elemental and isotopic abundances have become better constrained 
and compared to abundances in meteoritic material, presolar grains, and interstellar gas,
it has also become increasingly apparent that the Sun itself might have been contaminated by 
supernova material.  Surprisingly, large variations in elemental abundances among 
planet-hosting stars point to a similar stochastic contamination process by individual nearby supernovae.

The traditional environment articulated for this contamination  has been either ejecta sweeping 
over a distant  $\gtsimeq 10 \, {\rm pc}$ molecular cloud core, injecting material as it prompts
its collapse, or ejecta sweeping over a nearby ($\sim 0.1-1 \, {\rm pc}$)
protoplanetary disk. 
Here we consider for the first time enrichment in the H {\sc ii} region environment in which 
a core-collapse supernova is likely to take place. 
The explosion of a massive ($\gtsimeq 40 \, M_{\odot}$) progenitor will occur within
only 5 Myr, before the ionization fronts launched by the progenitor can advance
more than a few parsecs.  At these times
the material ejected by the supernova will interact with the molecular gas at
the edge of the H {\sc ii} region.

Supernova do not, in general, explode isotropically.   Instead,  both
numerical calculations and observations of SN1987A and the Cas A 
supernova remnant indicate that clumpiness is a common feature.
This clumpiness  plays a crucial role in enrichment, as
our numerical simulations find that isotropically exploding ejecta are too diffuse
to penetrate into a molecular cloud.
On the other hand, clumps with properties consistent with those in  Cas A deposit their material
$\approx 0.5$ pc into the molecular cloud, but only if cooling is significant, such thatthe cooling timescale is $\ltsimeq 10^2 \, {\rm yr}$.
Our simulations are limited by numerical resolution and were not able to span 
the entire set of relevant parameters, but these results appear  robust.

The gas at the edge of an H {\sc ii} region is widely recognized to be the site
of active star formation. It is likely that this star formation is triggered by the advance 
of the ionization fronts into the molecular gas, but the mechanism need not be identified to
assert that the supernova material injected into the molecular gas at the
edge of the H {\sc ii} region will be taken up by forming solar systems.
All of this star-forming material will be contaminated at
an average mixing ratio $\sim 10^{-4}$. 

Both this mixing ratio and the compositions of small regions within modeled core-collapse
supernovae are consistent with the quantities of ${}^{26}{\rm Al}$ injected
into the early solar system as well as the elemental and isotopic shifts inferred in
oxygen.  Possibly injection of ${}^{28}{\rm Si}$-rich silicon could also explain the difference 
in Si isotopes between the Sun and presolar grains.
Future work will examine whether specific regions within promptly exploding
supernovae match the isotopic shifts inferred from meteorites and other observations.

Injection of clumpy supernova material into molecular gas at  the edge of an H {\sc ii} region can occur
under very  common conditions, and all of the stars forming late in the evolution of an H {\sc ii} region are likely 
to  be contaminated by this process.  Depending on the specific trigger for star formation and the overall rate of star formation, 
we estimate between 7 and 30\% of {\it all} Sun-like stars are likely to be contaminated by a single, nearby supernova.
The injection process that we infer gave the solar system its inventory of ${}^{26}{\rm Al}$
and other isotopic anomalies may be a common, universal mechanism. 

\acknowledgements

We thank the referee, Alan Boss, for helpful comments and suggestions to improve the paper. We gratefully acknowledge 
support from the NASA Astrobiology Institute (08-NAI5-0018), from NASA Astrophysics Theory grant NNX09AD106 
to E.~S., and the National Science Foundation (grants AST 11-03608 to E.~S., and AST 09-07919 to S.~D.).  
We thank Mordecai-Mark Mac Low for helpful comments and Mark Richardson for help with the yt visualization package. 
All simulations were conducted at the ASU Advanced Computing Center, using the FLASH code, a 
product of the DOE ASC/Alliances-funded Center for Astrophysical Thermonuclear Flashes at the 
University of Chicago.

\end{document}